\documentclass{report}
\usepackage{setspace}
\usepackage{geometry}
\usepackage{graphicx} 

\usepackage{hyperref}
\hypersetup{colorlinks,allcolors=blue}

\hypersetup{
	bookmarksnumbered,		
	unicode,			
	}
\usepackage{mathtools}

\newcommand{\bigO}{\mathcal{O}}
\newcommand{\bra}{\langle}
\newcommand{\ket}{\rangle}

\begin{document}

\begin{titlepage}
\begin{center}

    \uppercase{Stony Brook University}\\
    \vspace{2mm}
    {Department of Physics and Astronomy}\\
    \vspace{2mm}
    {Department of Mathematics}

\vspace{5mm}

    \begin{figure}[h]
        \centering
        \includegraphics{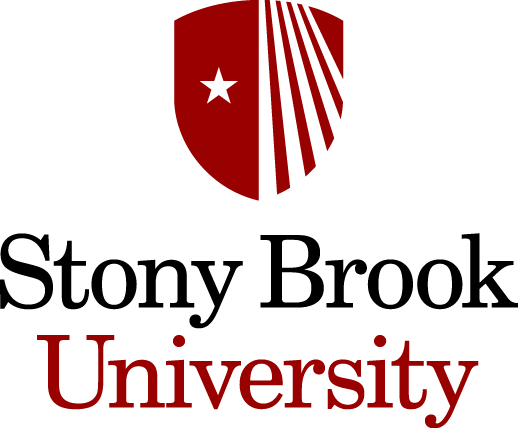}
    \end{figure}
    
\vspace{2cm}

    \begin{Large}{\textbf{Dynamics of Heavy Operators in $AdS_3$/$CFT_2$}}
    \end{Large}

    \vspace{1cm}
    
    \text{Aryaman Mishra}
    
    \vspace{0.5mm}
    
    \text{December 2023}
    
\vfill

    \begin{large}
    {\textbf{Honors Thesis to partially fulfill requirement for honors requirement in Mathematics and Physics}}
    \end{large}
    
\vspace{10mm}

    \begin{large}
 Global Advisor: Shiraz Minwalla \\
    Professor, Department of Theoretical Physics, Tata Institute for Fundamental Research, Mumbai, India \end{large}

\vspace{5mm}

    \begin{large}
    Local Advisor: Martin Roček \\ Distinguished Professor, C.N. Yang Institute for Theoretical Physics, Stony Brook University, NY
    \end{large}

\end{center}
\end{titlepage}

\newpage
\chapter*{\centering Abstract}

Correlation functions form the basis of Conformal Field Theories. These functions for two-point and three-point function have a restrictive closed form. In AdS/CFT the boundary of the geometry is where the CFT lives in d-dimensions, while the bulk is described by the AdS geometry living in d+1 dimensions. The correlation function in Ads/CFT are correlation of the operator insertions on the boundary (at CFT) through the complete geometry of bulk. These are represented by Witten diagrams \cite{Witten:1998qj} which at tree level doesn't have any quantum corrections. Generally, correlation functions are of low scaling (or conformal) dimension, $\Delta$, which is related to the mass of insertion of the scalar operator by, $\Delta(\Delta - 1) = m^2 L_{AdS}^2$. At low scaling dimensions the operator insertion on the CFT boundary does not back-react the metric of the geometry. On the other hand, at large scaling dimensions which scale with central charge the operator is considered heavy. These heavy operator were studied in by Maloney and Cardy \cite{Maloney:2019weq} \cite{Cardy:2017qhl}. This leads to an interesting question of what in the dual bulk (AdS) geometry of such heavy operators.  At the heavy limit $\Delta = m L_{AdS}$, which means that the mass of the operator insertion is large too. The two-point function of heavy-operator is assumed to be Black hole in $(d+1)$-dimensions which is studied in \cite{Vieira:2023jye} and the two-point form of CFT is recovered by calculating the action. In $3$-dimension we have more control over the geometry because of existence of exact metric \cite{Bañados:1999} with boundary stress-tensor insertion along with an exact map \cite{Roberts:2012aq} which maps it to Euclidean Poincare upper half plane. These methods are used in \cite{Vieira:2023bqv} to find the geometry for three-point function. The geometry is not simply of a black-hole but a wormhole solution for whose action is calculated which recovers the "square" of the classical DOZZ formula \cite{DORN1994375} \cite{ZAMOLODCHIKOV1996577}. 


We review the recent work of \cite{Vieira:2023jye} \cite{Vieira:2023bqv}  in this thesis to form an understanding of heavy operators in the context of AdS/CFT. The figures in this work are also sourced from them.

\thispagestyle{empty}
\clearpage

\newpage
\chapter*{\centering Acknowledgements}

As I complete this significant phase of my academic journey, I am overwhelmed with gratitude for the numerous mentors and educators who have guided, inspired, and supported me throughout my undergraduate years. Their dedication and expertise have been instrumental in shaping both my understanding and passion for physics.
Firstly, I extend my deepest appreciation to Prof. Radu Ionas, whose teachings in mechanics and topology went beyond academic instruction, providing me with a profound understanding of physics and invaluable life lessons. His mentorship has been a cornerstone of my academic development.
My heartfelt thanks to Prof. Warren Siegel, whose commitment and responsiveness to my endless inquiries have profoundly impacted my academic growth. His exceptional teaching, derived from his own publications, introduced me to the complexities and wonders of serious physics.
I am profoundly indebted to Prof. Peter Van Nieuwenhuizen for his exceptional courses in string theory, group theory, and advanced quantum field theory. His rigorous oral examinations and willingness to dedicate extra time for discussions not only deepened my understanding but also kept my joy for physics alive with fascinating olden tales.
The author express their sincere gratitude to Prof. Martin Rocek, who has been a constant source of support and guidance since my freshman year. His unique approach to teaching string theory and his role as a co-advisor in this work have been invaluable.
Special thanks are due to Prof. A. B. Zamolodchikov for introducing me to quantum field theory, statistical mechanics, and conformal field theory through his engaging and insightful course.
The author also extends his appreciation to all faculty members at Stony Brook University. Their comprehensive and challenging courses have significantly contributed to my foundational understanding of the subject.
Finally, I am deeply grateful to my undergraduate research advisor, Prof. Shiraz Minwalla, for the opportunity to engage in meaningful research at the scenic TIFR Mumbai campus. His mentorship, including his patience in handling communications across time zones, and his insightful course on conformal field theory, have greatly influenced my academic path. This work, inspired and directed under his guidance, is a testament to his ingenuity and mentor-ship.

I would also thank my friend Anirudh Deb from Stony Brook for discussion about this work and graduate students at TIFR especially Chintan Patel and Vineeth Krishna for extended discussions at TIFR.

\thispagestyle{empty}
\clearpage

\newpage
\tableofcontents

\newpage
\chapter{Introduction}

    The operators in the AdS/CFT generally talked about are light operators, these are computed using the Witten diagrams. 
    These operators scale with $\bigO(1)$. Witten diagrams are solutions to fields propagating through the bulk-boundary and their combinations. When the Witten diagram computation turns out to be cumbersome, we fall back on geodesic approximation of the propagators \cite{Louko_2000}. This is discussed in the two-point function chapter of the review.
    Heavier operators such as D-Branes scale as $\bigO(N)$. This kind of operator depends on the details of the CFT such as the D-branes for $\mathcal{N}=4$ Super Yang-Mills. The correlator of the D3 branes in $\mathcal{N}=4 \ SYM$ is related to the World volume of the D3 branes that approach the boundary in specified ways. For example the giant graviton configuration for D3 branes. \cite{McGreevy:2000cw}
    
    We are interested in very heavy operators which scale with $\bigO(N^2)$, these operators back-react on the metric of the geometry.These operators can be thought of as N-stacked Branes but we can generally write the Lagrangian for N Branes and has some analytic control. We have no Lagrangian description of these sort of operators. These contribute to the microstates of black holes. On the CFT side, these are defined by large conformal dimension $\Delta \sim L_{AdS}/G_N$. It is expected that these types of operators can be classified using modular invariance on (n-1)-genus surfaces \cite{Cardy:2017qhl} for n-point correlators. The only way to explore these operators is that we hope that they are described by black-hole physics on the gravity side. The black hole though is an ensemble physics and we have to work with free energy (entropy dependence) when dealing with it rather than energy (Mass term). We review the work in \cite{Vieira:2023bqv}\cite{Vieira:2023jye} where using clever techniques we calculate recover the two-point and three-point function of CFT while exploring the black holes. In this thesis, we explore this adventurous activity.


    \chapter{Two Point correlation function}

    \section{The Two-point function of light operators}

    We begin by reviewing the two-point function. These are light operator, the operators which do not bother the metric, insertions at the boundary, and are the tree-level Witten \cite{Witten:1998qj} diagrams. \cite{ammon2015gauge} \cite{Freedman:1998tz} 

    In d+1 dimensions the metric of Euclidean Ads metric is taken as 
    \begin{equation}\label{eq:1}
        ds^2=\frac{L^2}{z_0^2}(dz_0^2+\delta_{\mu \nu} dz^{\mu}d\bar{z}^{\nu})
    \end{equation}

    The toy action of super-gravity action relevant to us is 
    \begin{equation}\label{eq:2}
        S[\phi]=\frac{C}{2}\int dz_0 d^dz \sqrt{g} (g^{mn}\partial_m\phi \partial_n\phi+m^2\phi^2)
    \end{equation}
        Here, the relation of mass is $m^2L^2=\Delta(\Delta-d)$

    Taking the variation of this action, we get EOM:
    
    \begin{equation}\label{eq:3}
        (\frac{1}{\sqrt{g}}\partial_m\sqrt{g}g^{mn}\partial_n-m^2)\phi=0
    \end{equation}
    
    This is the Klein-Gordon equation. When inserting the metric Eq.\ref{eq:1} to the KG Equation \ref{eq:3} we find: 
    \begin{equation}\label{eq:4}
        z_0^{d+1} \frac{\partial}{\partial z_0} \left[ z_0^{-d+1} \frac{\partial}{\partial z_0} \phi(z_0, \vec{z})\right] + z_0^2 \frac{\partial}{\partial \vec{z}^2}
        \phi(z_0, \vec{z})- m^2 \phi(z_0, \vec{z})  = 0
    \end{equation}

    We observe that a solution which which becomes zero when $z_0 \rightarrow \infty$  and has a behavior of $\phi(z_0, {z}) \rightarrow z_0^{d - \Delta} \phi_{0}({z})$ as $z_0 \rightarrow 0$, where $\Delta = \Delta_{+}$ is the largest root where $\Delta_{\pm} = \frac{1}{2} (d \pm \sqrt{d^2 + 4 m^2} )$. Here $L_{Ads}=1$.

    A bulk-to-boundary green's function (propagator) was constructed in \cite{Witten:1998qj} which used the arguments of that it respects isometries of AdS space and hence is translational invariant and follows inversion.

    \begin{equation}\label{ads-propagator}
        K_{\Delta}(z_0,  z ,x) = 
        \frac{\Gamma(\Delta)}{\pi^{\frac{d}{2}} \Gamma(\Delta - 
        \frac{d}{2})} 
        \left( \frac{z_0}{z_0^2 + (z-x)^2} \right)^{\Delta}
    \end{equation}

    So the solution for source-free field is:

    \begin{equation}\label{solution-to-eom}
        \phi(z_0, {z}) = \frac{\Gamma(\Delta)}{\pi^{\frac{d}{2}} \Gamma(\Delta -\frac{d}{2})} \int_{\partial_{AdS}} d^{d} x\left( \frac{z_0}{z_0^2 + ({z}-{x})^2} \right)^{\Delta}\phi_0(\vec{x}) 
    \end{equation}

    and also \[\lim_{z_0 \rightarrow 0} z_0^{\Delta-d} K_{\Delta}(z_0, \vec z , \vec x) \rightarrow \delta^d(z - x)\]

    Now in Ads space we have inversion which leads the measure to transform like $d^{d}x  =   \frac{d^{d}x^{'}}{|{x}^{'}|^{2d}} $

    \[\left( \frac{z_0}{z_0^2 + (\vec{z}-\vec{x})^2} \right)^{\Delta} =
    \left( \frac{z'_0}{(z'_0)^2 + (\vec{z'}-\vec{x'})^2} \right)^{\Delta} |
     \vec{x} ^{'}|^{2 \Delta}\]

     We preserve the form of the shifted field at z:
     \[\phi(z')=\frac{\Gamma(\Delta)}{\pi^{\frac{d}{2}} \Gamma(\Delta -  \frac{d}{2})} 
    \int d^{d} x\left( \frac{z_0}{z_0^2 
    + (\vec{z}-\vec{x})^2} 
    \right)^{\Delta} \phi'_0(\vec{x})\]

    This shows that the scalar field transformed from $\phi'_0(\vec{x}) = |\vec{x}^{'}|^{2 (d-\Delta)} \phi_0(\vec{x'})$ . This leads to the idea that an operator inserted will transform as ${\cal O}(\vec{x}) \rightarrow {\cal O}'(\vec{x}) = |\vec{x'}|^{2 \Delta} {\cal O}(\vec{x'}) $ so that ${\cal O}(\vec{x})$ has dimension $\Delta$. Where operator insertion is viewed as $\int d^{d}x {\cal O}(\vec{x}) \phi_0(\vec{x})$.

    Now to evaluate the toy super-gravity action \ref{eq:2} we notice that we can integrate it by parts and only need to evaluate the boundary integral because $\phi$ by construction follows the equation of motion $\partial_m\partial^n \phi = m^2\phi$. By parts will loosely give us $\partial(\phi\partial\phi)=\partial\phi\partial\phi+\phi\partial^2\phi\ $ where the second part of equation will be replaced by EoM which will give us $\partial(\phi\partial\phi)=\partial\phi\partial\phi+m^2\phi^2$ So we can integrate only over the boundary and be done. We replace the scalar field $\phi$ with the full form in \ref{solution-to-eom}. These are bulk-to-boundary propagator solution with opposite end-points on the boundary meeting in the middle at a common point. The integral now becomes after integrating moving to boundary and taking $z_0=\epsilon$ as cut-off to regularize our integral.

    \[\label{2-point-function}
        \langle{\cal{O}}(\vec x) {\cal{O}}(\vec y)\rangle =
        C \,\lim_{\epsilon \rightarrow 0}
        \int \frac{d^dz}{z_0^{d-1}} K_{\Delta}(z_0, \vec z, \vec x)
        \left[ {\partial_{z_0}} K_{\Delta}(z_0, \vec z, \vec y)
        \right]_{z_0= \epsilon}
        \\
        = C \, \frac{ \Gamma[\Delta+1]}{ \pi^{\frac{d}{2}} \Gamma[\Delta- 
        \frac{d}{2}]}
        \frac{1}{|\vec x - \vec y|^{2 \Delta}}\]

    Where C is an arbitrary constant which will normalize our two-point function. Because of inversion property of the green's function and observing that the dimension of operator $\bigO=\Delta$ we get a scaling factor of $2\Delta$ and by nature of translation invariance of the propagator we see that the full two-point function is transnational invariant.

    \section{Correlation function from geodesic approximation} 

    Action of a relativistic particle through space is $S=-m \int ds$. Where $ds^2=g_{\mu\nu}\frac{dX^\mu}{d\lambda}\frac{dX^\nu}{d\lambda}$.
    Where the metric $g_{\mu\nu}$ is for the $AdS$ space. The particle propagating through the bulk is what is known as the "geodesic propagator" which is an approximation to the propagator or the Two-Point function. This is equivalent to the euclidean transition amplitude in Quantum Mechanics. Which becomes the path integral with the action that of geodesic. \cite{Louko_2000} 

    \begin{equation}\label{eq:5}
    \bra\phi(x)\phi(x')\ket = \int dP \ e^{-\Delta L(P)}
    \end{equation}

    Here $L(P)$ is the geodesic length through path $P$. $dP$ is the measure through all paths

    With the saddle point approximation\footnote{Also known as the steepest descent method is defined as $$\int d^n x e^{-S(x)/\lambda} = (2\pi\lambda)^n det(S)^{-1/2} e^{-S(x')/\lambda}$$ where $\nabla S|_{x=x'}=0$. Refer Sec 1.3 \cite{10.1093/oso/9780198834625.001.0001} } of the euclidean path integral we proceed to see that 
    \ref{eq:5} will become
    
    $$
    \bra\phi(x)\phi(x')\ket = \frac{\Gamma(\Delta)}{\Gamma(\Delta+1-d/2)} e^{-\Delta L(P)}
    $$

    In Euclidean AdS we take the metric with transformation of $r_i=\hat{\Omega_i}e^{\tau}$ and $z=e^\tau cos(\rho)$. As $z \rightarrow 0 $ we approach the boundary which in this new coordinate can be approached by $\rho \rightarrow \frac{\pi}{2}$. 

            $$
            ds^2= sec^2{\rho}(d\tau^2+ d\rho^2+ sin^2 \rho d\Omega^2) \text{ where } \Omega = \text{ volume metric }
            $$
        In the geometric-slice of cylinder where we only move in $\rho$ direction the metric will be $ds=sec(\rho)d\rho$ and hence the geodesic would be $\int_{-\rho}^{\rho} sec(\rho)d\rho$. For points on opposite side for simplicity
        the geodesic length is $2\log |Sec(\rho)+tan(\rho)|$. We approach the boundary $\rho=\pi/2-\epsilon e^{-\tau}$. In this limit geodesic is $2\log(\pi/2-\rho) = 2(\tau - \log \epsilon) $.
        We write without the normalization factor putting the geodesic in Eq. \ref{eq:5} \footnote{$ \bigO(t,\Omega) \equiv \lim_{\epsilon \rightarrow 0 } \epsilon^{-\Delta} \phi(t,\rho,\Omega) $ \cite{DiFrancesco:1997nk} which is relation between fields and operator. It is done as so to preserve conformal transformation property of $\phi$ which get inherited by CFT operator $\bigO$}.
        
        \begin{equation} \label{eq:6}
            \bra\bigO(x)\bigO(x')\ket = \lim_{\epsilon \rightarrow 0 }\frac{e^{-2\Delta(\tau-\\ln \epsilon)}}{\epsilon^{2\Delta}} = \frac{1}{e^{2\Delta \tau}}
        \end{equation}

        Define points on opposite side as mentioned earlier $x=\hat{\Omega_i}e^{\tau}$ and $x'=-\hat{\Omega_i}e^{\tau}$. We then recover the CFT two-point function as promised \cite{kaplan-no-date}
        \begin{equation}  \label{eq:aa}
             \bra\bigO(x)\bigO(x')\ket = \frac{1}{e^{2\Delta \tau}} = \frac{1}{|x-x'|^{2\Delta}}
        \end{equation}

    \section{Two-point correlation function of Heavy operator in AdS}

    A two-point correlation function of heavy operators needs to reproduce the same inverse of distance dependence. When we take the saddle point approximation of the action of the Black-hole instead of geodesic like in earlier section we should reproduce the same distance dependence as so it represent the actual bulk dual of the two-point function. We first introduce a geometry to visualize the calculation and proceed to calculate its action along with counter term. We then take a saddle point approximated propagator to reproduce the CFT two-point function. We will be putting down a counter-term and another term which comes from the "Stretched Horizon" to get the right two-point function scaling which otherwise would involve entropy because after-all a Black-hole is a thermal ensemble. 
    
        \subsection{Banana Geometry}

        We start with a Euclidean $(d+1)$-dimensional Ads Black-Hole. 

        \begin{equation}\label{eq:9}
            ds^2_{global}=f(r) d\tau^2+f(r)^{-1} dr^2+r^2d\Omega^2_{d-1}
        \end{equation}

        Where volume of sphere,
        
       \begin{equation} \label{eq:bb}
           \Omega_{d-1}=\frac{2\pi^{d/2}}{\Gamma(d/2)}
       \end{equation}
        
        and f(r) is the blackening factor:

        \begin{equation}\label{eq:10}
            f(r)=1+r^2-\frac{\alpha M}{r^{d-2}}
        \end{equation}
        
        Where M is the mass of Black-Hole and $\alpha$  is a parameter described by
        
        \begin{equation}\label{eq:11}
            \alpha = \frac{16\pi G_N}{(d-1)\Omega_{d-1}}    
        \end{equation}

        We set the Ads length, $L_{AdS}=1$, and $\alpha=1$ to get simpler calculations. We will introduce them again if needed. 

        We now will proceed do a change of coordinate from Global to Poincare such  that, $r=0$ and $-\infty \leq \tau \leq \infty$ is mapped to points of insertion in Poincare Ads. This transformation is called GtP (Global to Poincare):

        \begin{equation}\label{eq:12}
            \tau= \frac{1}{2}  \log(z'^2+R^2) \ ; \ r= \frac{R}{z'}
        \end{equation}

        This maps cylinders of different constant radius to cones of different radius anchored at origin.

        This leads to metric of Black-Hole to take the form:

        \begin{equation}\label{eq:13}
            ds^2_{cone} = \frac{1}{z'^2} \left[ \frac{dz'^2}{h(R/z')} + h(R/z') \left( dR+\frac{R}{z'}v(R/z')dz' \right)^2 + R^2 d\Omega^2_{d-1} \right]
        \end{equation}
         Where 
            $$h(r)=\frac{1}{f(r)}+\frac{r^2 f(r)}{(1+r^2)^2} \ ; \ v(r) = \frac{1}{f(r)h(r)} \left[ \frac{f(r)^2}{(1+r^2)^2}-1 \right] $$

        At $h=1$ and $v=0$ this metric becomes that of regular Poincare Ads.

        Now we have the geodesic running from origin to infinity in the cone metric. We want to pull the insertion at infinity to somewhere finite. To achieve this we do a special conformal transformation (SCT). The mapping pulls the geodesic end at infinity and puts it at at finite distance. 

        The SCT is described as seen in any CFT book as:

            \begin{equation} \label{eq:14}
                \centering
                x'^2 \longrightarrow x^i = \frac{x'^i-b^i\left(x'^2+z^2\right)}{1-2b.x+b^2(x'^2+z'^2)}
                \text{ and }
                z'\longrightarrow z = \frac{z'}{1-2b.x+b^2(x'^2+z'^2)} 
            \end{equation}

        The special conformal transformation is an inversion 
        $\left(x^\mu \longrightarrow \frac{x^\mu}{x^2} \right)$ 
        followed by a translation 
        $\left(\frac{x^\mu}{x^2} \longrightarrow \frac{x^\mu}{x^2}-b^\mu \right)$ 
        and then another inversion 
        $\left(\frac{x^\mu}{x^2}-b^\mu \longrightarrow \frac{1}{\frac{x^\mu}{x^2}-b^\mu} \right)$.
        
        The SCT mapping pulls the geodesic end originating from origin and ending at infinite to a near finite distance. When this idea applied to cone can be visualized as such. The far end of the cone gets pulled to a finite point and gets pinched. So both ends are cones. This geometry looks like a banana which is referred to as "space-time banana geometry". 
        
        The metric of this geometry is:
            \begin{equation}\label{eq:15}
                ds^2_{banana} = N^2_z dz^2 + \sum h_{ab}(dy^a+N^a_z dz)(dy^b+N^b_z)+\frac{\rho^2 d\Omega^2_{d-2}}{z^2}
            \end{equation}
        
        The metric of geometry itself is not enlightening and not useful but how this transformation leads us to new geometry will help us.

        \begin{figure}
            \centering
            \includegraphics[scale=0.4]{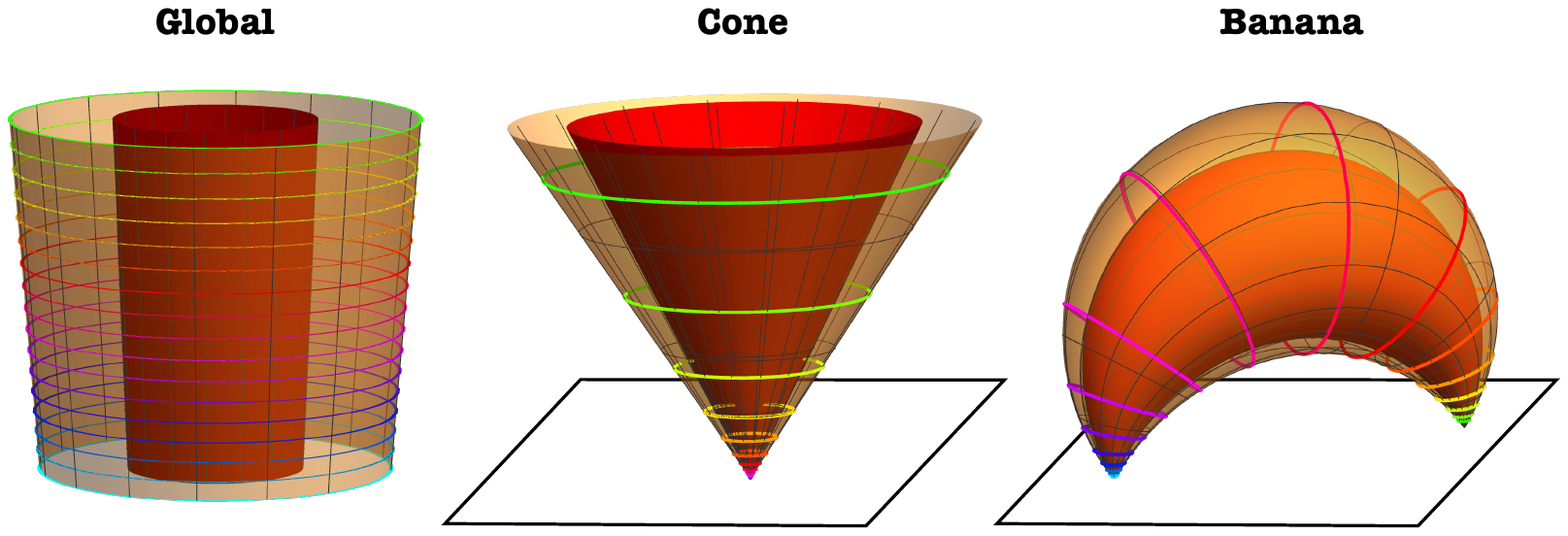}
            \caption{The image illustrates the equivalent Black Hole in all three different geometries via the transformation. The left is AdS global followed by cone and banana (right)}
            \label{fig:global_to_cone_to_banana}
        \end{figure}

            When $b=0$ we have foliations by cone when b becomes a finite value we get foliations by banana. Where each foliation is surface of constant r. The mapping from global to banana is 
            \begin{equation}\label{eq:16}
                r^2=\Theta^2 \left(\frac{x^2}{z^2}+1\right)-1 \  ; \ \Theta=1+2b.x+b^2(x^2+z^2) 
            \end{equation}
            We can algebraically solve for $z$ and see that we get two positive solutions because of square root. The larger root $z_+$ is the cap of the banana (the upper half) while the smaller root $z_-$ is pair of pants pinched at the bottom. When $b=0$ the upper half goes to infinity and the pinched pants becomes cone as mentioned earlier. 

        \subsection{The on-shell action from gravity}

        To see if this is the gravity dual of CFT two-point function we'll have to find the on-shell action and calculate the propagator. We'll use the same saddle-point approximation of the propagator to see if it matches.
        
        \begin{equation}\label{eq:17}
            \langle \bigO(x_1)\bigO(x_2) \rangle = e^{-S} \sim \frac{1}{|x_1-x_2|^{2\Delta}}
        \end{equation}

        We should hence expect the behavior of the action to be 

        \begin{equation}\label{eq:18}
            S = 2\Delta|x_1-x_2|+ \text{distance-unrelated constant}
        \end{equation}

        The total bulk action of two point function is

            \begin{equation}\label{eq:19}
                S = S_{bulk}+ S_{boundary} + S_{counter-term}
            \end{equation}

        The boundary action is divided in two Gibbons-Hawking-York terms
        
            \begin{equation}\label{eq:20}
                S_{boundary}    = S_{GHY}\left(\partial_{AdS}\right)+S_{GHY}\left(\partial_{\text{stretched-horizon}}\right)
            \end{equation}

            \subsubsection{The bulk action}

            The Einstein-Hilbert action for bulk is 

                \begin{equation}\label{eq:21}
                    S_{bulk}=-\frac{1}{16 \pi G_N}\int d^dx dz \sqrt{g} \left(R+\frac{d(d-1)}{L^2_{AdS}}\right)
                \end{equation}

            In d-dim Ads space $R_{\mu\nu}=\frac{-d+1}{L_{Ads}} g_{\mu \nu}$, here we choose $L_{AdS}=1$ as said earlier and the we consider (d+1) dim Ads so $R_{\mu\nu}=-d g_{\mu \nu}$. $R=g^{\mu\nu}R_{\mu\nu}$ which becomes $R=-d^2$. The metric component $\sqrt{g}=\frac{1}{z^{d+1}}$ same as empty Ads because of the SCT transformation.
            The z-integration will give us a $z^{-d}/-d$ component and the denominator will get cancelled by the $-d$ of the constant in integral. 

            After putting it all together we get following after $dz$ integration in domain $z\geq \epsilon$. The $\epsilon$ is a cut-off surface which will be crucial to regulate the action and the get the distance dependence.

            \begin{equation}\label{eq:22}
                8\pi G_N \ S_{bulk} = \int_{A} d^d x \left[\frac{1}{z_+^d(r_h)}-\frac{1}{z_-^d(r_h)}\right]+\int_{A\cup B}d^d x \frac{1}{\epsilon^d}
            \end{equation}

\begin{figure}
                \centering
                \includegraphics[scale=0.4]{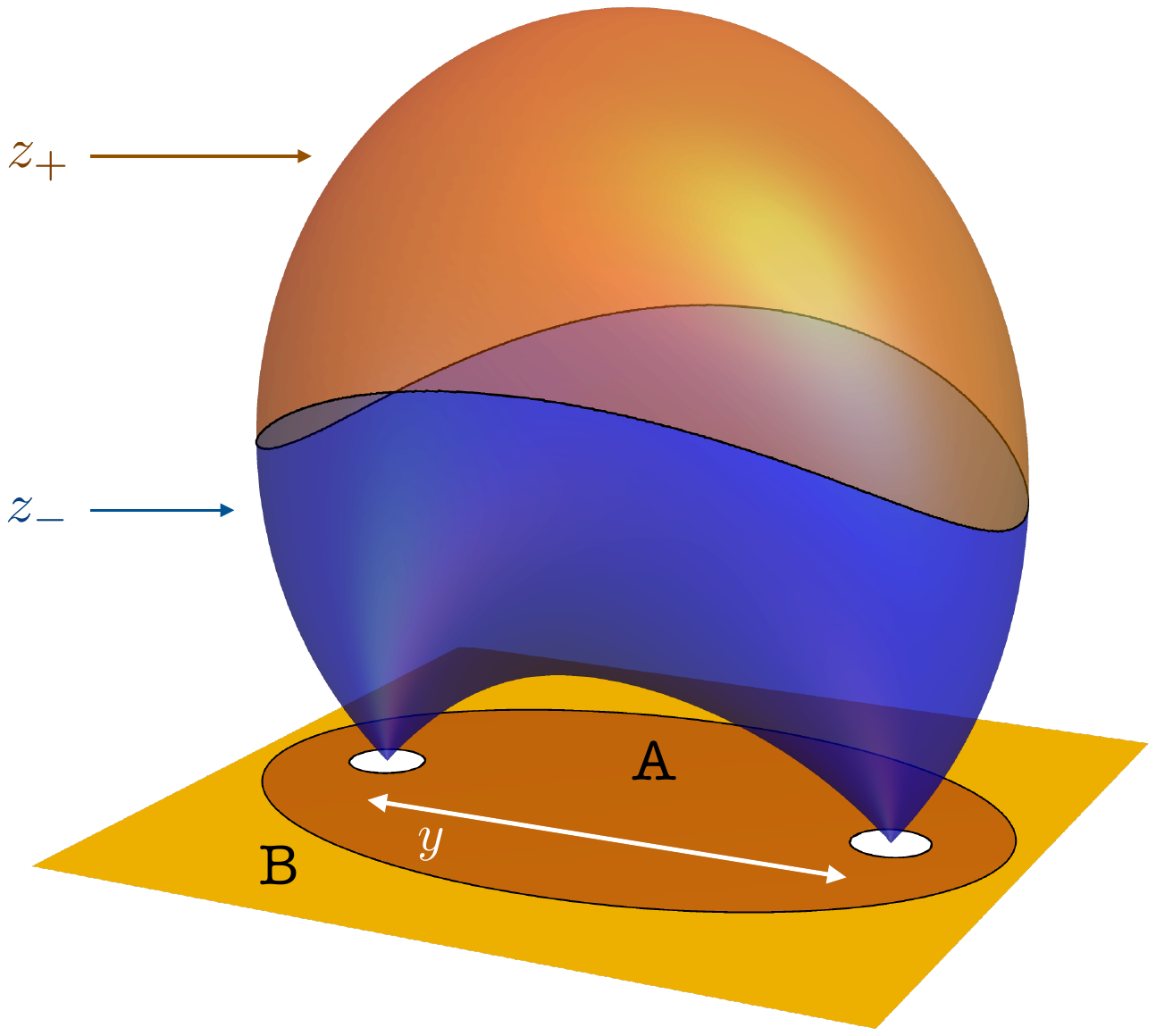}
                \caption{This figure shows the Banana geometry with the upper cap describe by $z_+$ and the lower pinched pants by $z_-$. There is also the domain of integration A and B which is inside and outside the projection of horizon respectively. The distance between the insertions is $y=\frac{1}{b}$.}
                \label{fig:Bananageometry}
            \end{figure}

                A is the domain of integration which is projection of banana as a boundary condition of horizon. It has a shape of ellipse as shown in Figure. B is the outside domain which has the divergent term because of AdS volume. This domain has been described in Figure \ref{fig:Bananageometry}

            When substitute $r_h$ in $z_{\pm}$ we get
            
            \begin{equation}\label{eq:23}
                z_{\pm}(r_h)= \left[ \frac{r_h^2}{2b^2}-\left(x^1+\frac{1}{b}\right)-\rho^2 \pm \frac{r_h}{b}\sqrt{\frac{r_h}{4b^2}-x^1(x^1+\frac{1}{b})-\rho^2\left(1+\frac{1}{r_h^2}\right)} \right]^{1/2}
            \end{equation}
            
            The domain A is described the the inequality which is locus of the above square root 

                    \begin{equation}
                    x^1(x^1+1/b)+\rho^2(1+1/r_h^2)\leq r_h^2/4b^2
                    \end{equation}
            
            We choose insertion on $x^1$ axis at $x^1=0 \text{ and } x^1=-1/b$.

            We'll now see how $z_-$ leads to point of insertion as in why we thought it is the pinched pants. We do a small $\delta$ expansion of $1/z_-$\footnote{Mathematica} near the insertion point of $x^1=0 \text{ and } x^1=-1/b$.
            
            \begin{equation} \label{eq:24}
                \frac{1}{z_-(r_h)}=\frac{r_h^d}{((x^1)^2+\rho^2)^{d/2}((1+bx)^2+b^2\rho^2)^{d/2}} + D(b)
            \end{equation}
        
            The function D is a function which is difference function but independent of the distance between the insertion points which becomes zero when b=0 i.e $D(0)=0$.
            The first term without the $r_h$ is the $f_{2pt}$ term.

            \begin{equation} \label{eq:25}
                f_{2pt}= \frac{|x_1-x_2|^d}{|x-x_1|^d|x-x_2|^d}
            \end{equation}

            This can be directly seen from \ref{eq:24} putting $\rho=0$ with $x^1=0 \text{ and } x^2=-1/b$. This is a function of just the distance. Something we're looking for.

            We now proceed back to our integral after describing the domains and geometry well. The $f_{2pt}$ is logarithmically divergent in domain A because of $\int d\rho/\rho$ dependence in $\int f_{2pt}$ term. We so add and subtract $f_{2pt}$ in B domain because it is integrable at infinity.

            \begin{equation} \label{eq:26}
                8\pi G_N \ S_{bulk} = \int_{A\cup B}d^d x \left[\frac{1}{\epsilon^d}-r_h^d f_{2pt}\right] + I_{bulk}
            \end{equation}

            $I_{bulk}$\footnote{$$8\pi G_N I_{bulk}= \int_{A} d^d x \left[\frac{1}{z_+^d(r_h)}-D(b)\right] + \int_{B} d^d x r_h^d f_{2pt}$$} is a constant independent of the distance between the insertion points. 

        \subsubsection{The boundary action}
        
        The boundary action of AdS boundary is the Gibbons-Hawking-York term.

            \begin{equation} \label{eq:27}
                8\pi G_N S_{GHY}= \int_{\partial} d^d x \sqrt{h} K_{\partial_AdS}
            \end{equation}

            where $h_{ab}$ is the boundary metric at $z=\epsilon$ cutoff surface and, \ref{eq:15}
            
            \begin{equation}
                K_{\partial_AdS}=-\nabla \frac{N^\mu_z}{|N_z|}
            \end{equation}
            is the extrinsic curvature. $N^\mu_z$ is the normal-vector pointing out from banana which was also there in the metric.

            The counter-term action for AdS boundary is sourced from \cite{Emparan:1999pm} Eq.17. 
            \begin{equation} \label{eq:27.1}
                 8\pi G_N S_{ct}=\int_{\partial} d^d x \sqrt{h} \left[(d-1)+\frac{R[h]}{2(d-2)}+\frac{1}{2(d-4)(d-2)}\left(R_{ab}[h]R^{ab}[h]-\frac{d}{4(d-1)}R[h]^2\right)+ \dots \right]
            \end{equation}

            Here we now mention a subtlety that would help us make series expansion of the boundary stress-tensor. We choose disks of radius "a" around the insertion points. In this approximation when $|x_1-x_2|>>a$ we have a series expansion of $S_{GHY}$ and $S_{ct}$ in terms of $\frac{z}{|x-x_1||x-x_2|}$ something which can be seen from \ref{eq:24}. This is because of the series expansion of SCT map which can be checked in mathematica.
            \begin{equation} \label{eq:28}
                \sqrt{h}= \frac{1}{z^d} \left[ 1-\frac{(\alpha M/2)z^d}{((x^1)^2+\rho^2)^{d/2}((1+bx)^2+b^2\rho^2)^{d/2}}+\dots \right]
            \end{equation}

            The expansion of extrinsic curvature is $K=-d+\bigO(z^{d+1})$ in the first order. Since extrinsic curvature is $K = -\nabla_\mu \frac{N^\mu_z}{|N_z|}$ and $-\nabla_i V^i = \frac{1}{\sqrt{g}}\partial_i(\sqrt{g}V^i)$. Where $\sqrt{g}=\frac{1}{z^{d+1}}$ which is same as empty AdS. Now when we expand the extrinsic curvature in first order it is just derivative of the metric term which is just "d" and negative sign gives us  $K=-d+\bigO(z^{d+1})$. The higher terms depend on $z^{d+1}$ because of the nature of the metric.

            \vspace{1cm}

            After combining the bulk, boundary and counter-term and putting limits on $z\geq\epsilon$ we get:
            
            \begin{equation}
                8 \pi G_N S_{bulk+boundary+ct}=\int d^d x \left[\frac{1}{\epsilon^d}-r_h^d f_{2pt}\right] + \sqrt{h}\left(-d+(d-1)\right) + \bigO\left(z^{d+1}\right)
            \end{equation}
            
            Where last term is combination of extrinsic curvature and counter-term in first order. Now we expand the $\sqrt{h}$ as mentioned in \ref{eq:28} and get a combined result of
            
            \begin{equation}   
                8 \pi G_N S_{bulk+boundary+ct}=\int d^d x \left[\frac{1}{\epsilon^d}-r_h^d f_{2pt}\right] - \left[\frac{1}{\epsilon^d}-\frac{\alpha M}{2} f_{2pt}\right]
            \end{equation}
            
                The final result for which is
                
            \begin{equation} \label{eq:29}
               S_{bulk+boundary+ct}= \frac{\alpha M - 2r_h^d}{16 \pi G_N} \int_{A\cup B- disc} d^dx f_{2pt}
            \end{equation}

            This domain is what we'll use unless otherwise mentioned. In this domain the integrand remains integrable it leads to no logarithmic divergence or as such and there is no blow-up near insertion point because we remove the "a" radii discs from domain around the insertion points. 

            The integral of $f_{2pt}$ will give us a logarithmic answer as discussed before.\footnote{This is solved in bi-polar coordinates because of two operator insertions. Appendix. We also do a Mathematica calculation in 3-dimension just to see the behaviour.} 
            The domain of integral has a cut-off at radius "a" to not include disc insertions. We change the integration variable from $\int d^dx \rightarrow \int \rho^{d-2}dx^1 d\Omega_{d-2}$.The integral over $d\Omega_{d-2}$ just gives volume of the sphere $\Omega_{d-1}$.  \footnote{$x_1 \rightarrow \frac{1+T cos \sigma}{b(1+T^2+2T cos \sigma}$ $\rho \rightarrow \frac{T sin \sigma}{b(1+T^2+2T cos \sigma}$ are the bipolar transformation which gives exact result}

            \begin{equation} \label{eq:30}
                \int \rho^{d-2}dx^1 d\Omega_{d-2} f_{2pt}= \Omega_{d-1} \log \frac{|x_1-x_2|^2}{a^2}
            \end{equation}

            The over result for $S$ upto now turns out to be
            \begin{equation}\label{eq:31}
                S_{bulk+boundary+ct} = \frac{\Omega_{d-1} (\alpha M - 2r_h^d)}{16 \pi G_N} \log \frac{|x_1-x_2|^2}{a^2}
            \end{equation}

            The Black-hole temperature and entropy are  
            \begin{equation} \label{eq:32}
                T_H = \frac{f'(r_h)}{4 G_N} \text{ and } S_{BH} = \frac{A}{4 G_N}
            \end{equation}

            Which are called as Bekenstein-Hawking temperature and entropy \cite{PhysRevD.7.2333} where $f(r)$ is the blackening factor \ref{eq:10}. The area of black hole horizon is  $A=r^{d-1}\Omega_{d-1}$. Then 
            
            \begin{equation}
                S_{BH}=\frac{r^{d-1}\Omega_{d-1}}{4 G_N}
            \end{equation}
            and
            \begin{equation}
                T_H=\frac{2 r_h}{4 \pi}+\frac{(d-1) \alpha M}{4 \pi r_h^{d-1}}
            \end{equation}

            We can see that then $$ST_H = \frac{r_h^d \Omega_{d-1}}{16 \pi G_N} + \frac{(d-2)\Omega_{d-1}}{16 \pi G_N} $$ 

            If we put it together we notice that 
            \begin{equation} \label{eq:33}
                F_{Gibbs} = M - ST_H \footnote{E=M by mass-energy equivalence} = \frac{\Omega_{d-1} (\alpha M - 2r_h^d)}{16 \pi G_N}
            \end{equation}

            This says that \ref{eq:31}
            
            \begin{equation} \label{eq:34}
                S_{bulk+boundary+ct} = F_{Gibbs} \log \frac{|x_1-x_2|^2}{a^2}
            \end{equation}

            This gives us a conundrum because we want mass dependence and not a dependence on entropy. This is because a CFT two-point function scales with the conformal dimension which by state-operator map corresponds to energy (= mass) not Gibbs free energy. To resolve this we introduce a new action of Black-Hole Stretched horizon. Which we evaluate now and when added to the action will remove the $ST_H$ dependence and give the scaling solely on the mass of the operator insertions. 

            \subsubsection{The Stretched Horizon action}

            The stretched horizon is a membrane like structure which floats just above the horizon \cite{PhysRevD.58.064011}. The membrane picture is interpreted as added surface term of electromagnetic and gravitational sources residing on the stretched horizon. This is such because the stretched horizon unlike real horizon has no degeneracy in the induced metric and the surfaces are causal. This gives us a mathematical object which can describe objects in classical gravity and yet capture quantum effects just above the horizon.\cite{unknown-author-no-date} \cite{PhysRevD.33.915}. One of the reasons to assume it becomes relevant here is because the operator insertions in banana geometry pierces the boundary which along with the boundary of empty AdS makes the boundary of banana relevant. This boundary can be interpreted as the streched horizon. \footnote{\href{https://youtu.be/8V-9IInpzBM?si=-I6FgPuNcftz59Mf}{Explained in more detail in talk here}}

            The action for Stretched horizon is describes as a boundary term at $r=r_h(1+\epsilon`)$. This $\epsilon`$ is different from $\epsilon$ of cut-off surface
            We now evaluate this action.

            \begin{equation} \label{eq:34.1}
                8\pi G_N S_{GHY-Streched-Horizon}   = \int_{\partial} \sqrt{h_{stretch}} \ K_{stretch} \text{ in limit $\epsilon` \rightarrow 0 $ }
            \end{equation}

            $$K_{Stretch}= - \nabla_\mu \frac{V^\mu}{|V|} \text{where V is orthogonal to horizon like in asymptotic boundary case}$$

            Because $V_{\mu}$ is the orthogonal to horizon which will be the $r(x,z)$ \ref{eq:16} term. We'll need to track how r gets mapped into banana metric after-all because we're not in the global coordinate.  We consider the image through Poincare co-ordinate map and special transformation map of $\partial_r$.\footnote{$\frac{\partial}{\partial r} \rightarrow \frac{\partial r}{\partial x_\mu} \frac{\partial}{\partial r} $}.
            \footnote{where $\nabla_{\mu} V^\mu = \frac{1}{\sqrt{g}} \partial_\mu \sqrt{g} V^\mu$ and $\sqrt{g}=\frac{1}{z^{d+1}}$. It is a simple exercises from here}

            So the extrinsic curvature leads to be

            \begin{equation} 
                K_{Stretch}= \nabla_{\mu}\frac{\partial_{\nu} r(x,z)}{\sqrt{f(r)}} 
            \end{equation}


            which when evaluated at $r_h$ gives except the denominator. $f(r_h)=0$ and $f'(r_h)=4 \pi T_H$            
                 \begin{equation} \label{eq:36}
                K_{Stretch}= \frac{2 \pi T_H}{\sqrt{f(r)}} 
                \end{equation} 

            The determinant of induced metric is 

            \begin{equation} \label{eq:37}
                \sqrt{h_{stretch}} = \frac{\sqrt{f(r)}}{z^d r} \left(1+D`(b)\right)
            \end{equation} 

            Where $D`(b)$ is another difference such that $D`(0)=0$.

            We go ahead to evaluate the integral $\int \sqrt{h_{stretch}} K_{Stretch}$ at $r = r_h$ and $z = z_-$ because we're interested in the insertion point. The $z=z+$ part will be absorbed in extra constant.

                \begin{equation}
                    \int \sqrt{h_{stretch}} \ K_{Stretch} = 2 \pi r_h^{d-1} T_H \int d^dx f_{2pt}
                \end{equation}
                
                where Eq.\ref{eq:24} is used. The integral of $f_{2pt}$ gives a volume of the surface which with $r_h^{d-1}=\frac{4G_N S}{Vol}$ combined with external factors gives the entropy, $S$.

                \begin{equation} \label{eq:38}
                    S_{GHY-Stretched} = ST_H \log \frac{|x_1-x_2|^2}{a^2} + N_s
                \end{equation}

            \vspace{2cm}
            
            \subsubsection{The complete action and two point revelation}
                
                After we combine all the terms in actions we get \footnote{boundary = $\partial_{AdS}+\partial_{Stretch}$} \footnote{ $N = N_{bulk}+N_{s}+ -2 \log a$}

                \begin{equation}
                    S_{bulk+boundary+ct} = M \log \frac{|x_1-x_2|^2}{a^2} + N_{bulk} + N_s
                \end{equation}

                \begin{equation} \label{eq:39}
                    S_{bulk+boundary+ct} = 2 M \log |x_1-x_2| + N
                \end{equation}

                So when doing the saddle point approximation of the propagator we get \footnote{$\Delta = M$ for large $\Delta$ and the normalization constant has been omitted}

                \begin{equation} \label{eq:40}
                    e^{-S_{bulk+boundary+ct}} = \frac{1}{|x_1-x_2|^{2\Delta}}
                \end{equation}

    \section{ Two-point Heavy operator in three dimension in geometry of "doors and domes" }

            We have now established a space-time banana geometry where with right regularization and renormalization in arbitrary d-dimensions we can find an action whose approximated propagator behaves like a CFT two-point function. We will now go ahead to 3-dimensions. In 3 dimensions the Fefferman-Graham (FG) expansion of metric truncates at finite order. The resulting metric is known as Bañados metric.\cite{Bañados:1999} . We will now explore recreation of two point function in this geometry. 

            The back-reaction of dual operators induces an expectation value of boundary stress-tensor\cite{Balasubramanian:1999re} which is interpreted as

                \begin{equation} \label{eq:41}
                    \langle T_{ij} (x) \rangle = \frac{ \left\langle T_{ij} (x) \prod_{k=1}^n \bigO_{\Delta_k} (x_k)\right \rangle }{\left\langle \prod_{k=1}^n \bigO_{\Delta_k} (x_k) \right \rangle }
                \end{equation}

                This is the boundary expectation value to stress tensor of heavy operator insertions in correlation function. We'll explore the stress tensor on boundary in d=2 CFT. We'll first study two operator insertions which will fix n = 2. Later when studying three point function we'll fix n=3 and explore that aspect.

                The conformal dimension is large which sets \(\Delta \approx \frac{L^{d-1}}{G_N}\). We parameterize it $\Delta_i = \frac{c}{12} M_i$ where M is the mass of insertion. Here $c=\frac{3 L}{2 G_N}$. Where L is the AdS radius and $G_N$ is newton's gravitational constant. The parametrization is such that $M > 1$ is a black hole while $M < 1$ is a conical defect solution. The $M = 1$ is the point of phase transition between the two.

            \subsubsection{The Bañados metric}

            In three-dimensional gravity the FG expansion of metric sources from boundary stress tensor truncates at a finite order resulting in the metric \cite{Bañados:1999}

            \begin{equation} \label{eq:42}
                ds^2 = \frac{dy^2+dz d\bar{z}}{y^2} + L(z) dz^2 + \bar{L}(\bar{z}) d\bar{z}^2 + y^2 L(z) \bar{L}(\bar{z}) dz d\bar{z}
            \end{equation}
             
             Where \(\langle T_{zz} (z)\rangle = -\frac{c}{6} L(z)\) and same for the complex conjugate. The reason that the truncation is finite is because the three-dimensional gravity is topological in nature and doesn't have dynamical terms.

             \subsubsection{The Roberts Map}

             We now introduce the second essential component of out analysis. The Roberts map \cite{Roberts:2012aq}.  \footnote{This metric has a degeneracy at \(det (g) =0\) where it turns out to be \(y^{-4} = L(z) \bar{L}(\bar{z})\) which is the degeneracy surface. The newly formed geometry reveals beyond this degeneracy as well.}

             The map is defined as 
                 \begin{align} \label{eq:43}
                     Y = y \frac{4 \partial f \bar{\partial} \bar{f}^{3/2}}{4 \partial f \bar{\partial} \bar{f} + y^2 \partial^2 f \bar{\partial}^2 \bar{f}}
                     \qquad
                     Z = f(z) - \frac{2 y^2 (\partial f)^2 \bar{\partial}^2 \bar{f}}{4 \partial f \bar{\partial} \bar{f} + y^2 \partial^2 f \bar{\partial}^2 \bar{f}}
                     \qquad
                     \bar{Z} = \bar{f}(\bar{z}) - \frac{2 y^2 (\bar{\partial} \bar{f})^2 \partial^2 f}{4 \partial f \bar{\partial} \bar{f} + y^2 \partial^2 f \bar{\partial}^2 \bar{f}}
                 \end{align}

                 This mapping is special because it transforms the Bañados metric \ref{eq:42} to the Euclidean Poincare $Ads_3$ upper half plane. \footnote{At conformal boundary $y=0$, $Z = f(z)$ and $\bar{Z} = \bar{f}(\bar{z})$ where $ f \text{ and } \bar{f} $ are family of solutions related by mobius transformations such that \[f(z)= \frac{a f_0 + b}{c f_0 + d}\] }

                 \begin{align*}
                     ds^2 = \frac{dY^2 + dZ d\bar{Z}}{Y^2}
                 \end{align*}

                 Where the function $f(z)$ and $\bar{f}(\bar{z})$ are related to boundary stress tensor \footnote{$\{f,z\}$ is the schwarzian derivative defined as 
                 \[\{f,z\}=\frac{f'''}{f'}-\frac{3}{2}(\frac{f"}{f'})^2\]}

                 \begin{equation}\label{eq:schwarzian}
                 -\frac{1}{2} \{f,z\} = L(z) \text{ \ and \ } -\frac{1}{2} \{\bar f,\bar z \} = \bar L(\bar z) 
                 \end{equation}

                 \begin{figure}
                     \centering
                     \includegraphics[scale=0.4]{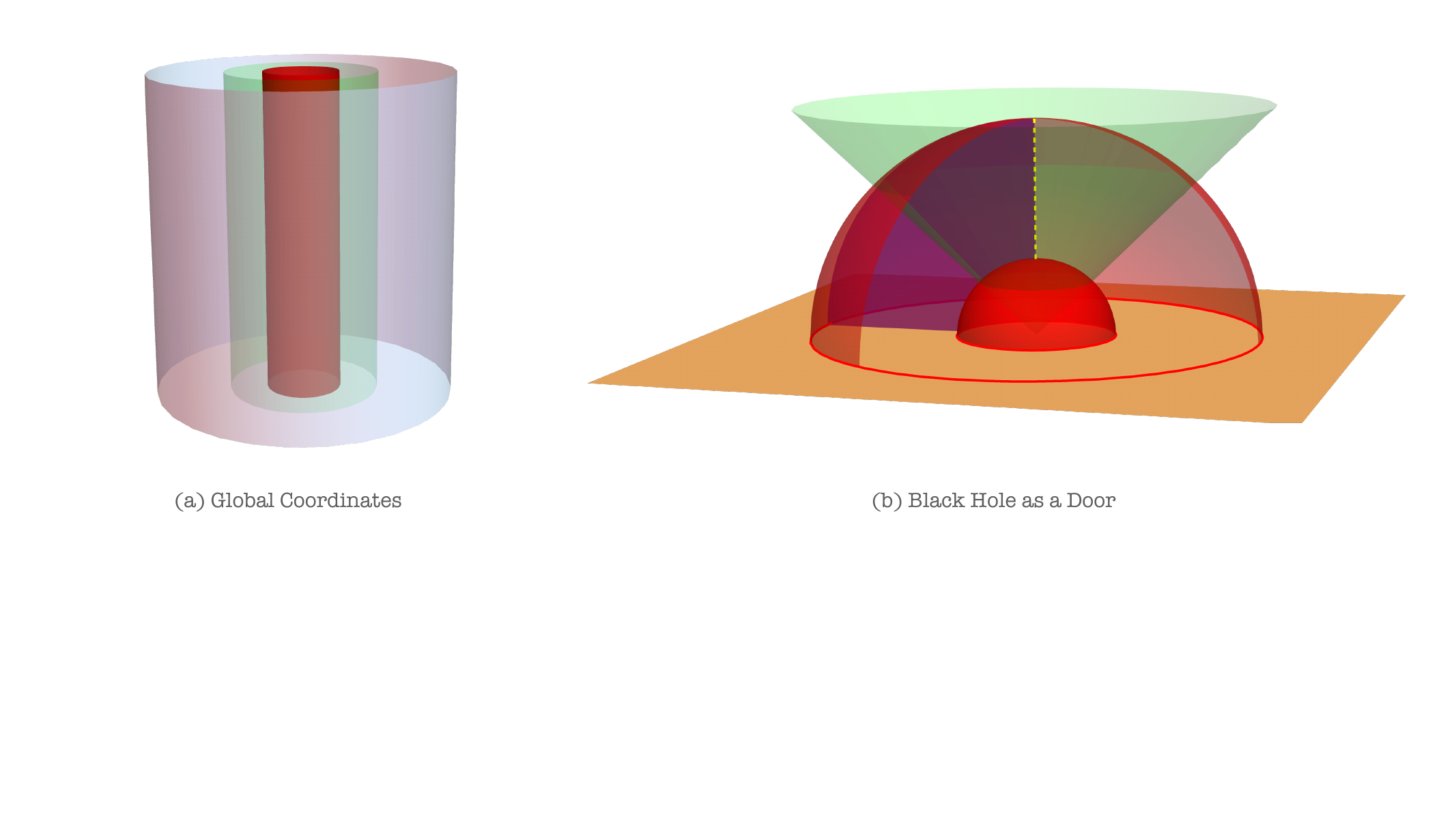}
                     \caption{We see the image of BTZ Black hole in Global co-ordinate and the equivalent geometry in the Dome. The red cylinder in middle is the horizon which in Dome picture is the smaller dome. The green cylinder in the global co-ordinate represents the degeneracy of metric which in dome is mapped as a cone.}
                     \label{fig:Black-hole dome with wall}
                 \end{figure}

            \subsubsection{Two-point function in door dome}

                 We invoke the \ref{eq:41} for the case of two point function. Where we get

                 \[ \langle T_{ij} (x) \rangle = \frac{\langle T_{ij} (x) \bigO_\Delta \bigO_\Delta \rangle}{\langle  \bigO_\Delta \bigO_\Delta \rangle}  \]

                 which leads us to   \footnote{Here we use the conformal ward identity for the numerator in the stress tensor expectation value and denominator is the regular two point function in CFT.} \footnote{we take operators to be spinless. Conformal spin is defined as $s=h-\bar{h}$, for spinless objects $h=\bar{h}$. Using the same definition we would like to remind the reader that in 2D, conformal dimension $\Delta = h+\bar{h}$. $h$ and $\bar{h}$ are called holomorphic and anti-holomorphic dimensions.}

                 \[L(z) = -\frac{M (z_1 - z_2)}{4 (z-z_1) (z-z_2)}\]  
                 
                 We now have to study the Banados metric with the $L(z)$ present. But we see that with roberts map the conformal boundary is mapped such that $Z=f(z)$. We set the insertion point at $z_1=0$ and $z_2= \infty$. This leads to \(L(z) = -\frac{M}{4 z^2}\). When solving the schwarzian equation for this \footnote{When we take schwarzian derivative of this equation we reproduce \(L(z) = -\frac{M}{4 z^2}\) }
                 
                 we get that \[f_{BH} (z) = z^{i \sqrt{M-1}}\] where we define $R_h=\sqrt{M-1}$. We have in the new mapping  at conformal boundary  $Z = f_{BH}(z)$. 

                 \begin{equation}\label{eq:44}
                     z = e^{\tau + i \phi} \longrightarrow Z = e^{R_h (i \tau - \phi)}
                 \end{equation}

                 So under the Roberts map the time and angle switch places as being real or imaginary. Our time ran from $-\infty to +\infty$ while $0 \leq \phi \leq 2\pi$. Now that they have switched places the rotation in "z-plane" will translate to dilatation in "Z-plane" and vice-versa. This leads us to introducing a crucial step in our analysis. We add a "branch cut" because the time-axis will cover across all real numbers while the annulus only runs from $0$ to $2\pi$.

                 \begin{figure}
                     \centering
                     \includegraphics[width=\textwidth]{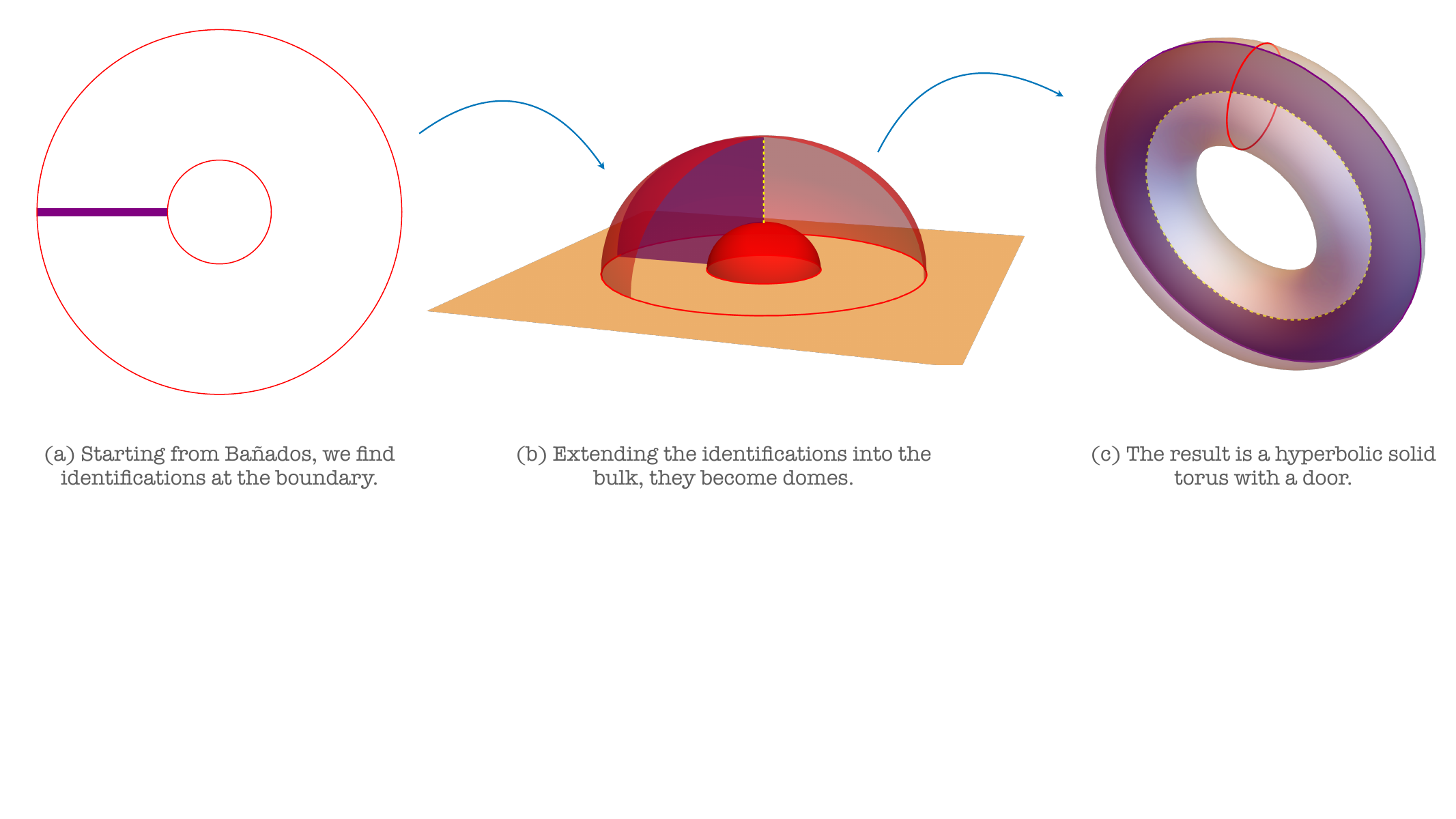}
                     \caption{2-point correlation function is represented as a dome with a branch cut. The inner dome is identified with outer dome which is equivalent to a torus as shown. This representation resembles schottky genus-1 identification \cite{krasnov2000holography} but with a added branch cut}
                     \label{fig:2ptdome}
                 \end{figure}

                The dome with branch cut extended into the bulk is the geometry which describes the two point function. This dome with branch cut is equivalent to the Banana geometry described above in 3-dimensions and is equivalent by existence of exact metric dependent on boundary-stress tensor and a mapping which maps it to empty AdS upper half plane. 

                The connection of north pole of both domes is where the black-hole horizon is situated. This is how in this representation of identification is the black-hole horizon represented.

                Finally, we do a calculation of the action to show that this setup recovers our regular CFT two-point function in Appendix. This needs the knowledge of liouville fields which we discuss extensively in three-point function section.

            \subsubsection{Isometry of two-point geometry}
            
                \begin{figure}
                    \centering
                    \includegraphics[width=\textwidth]{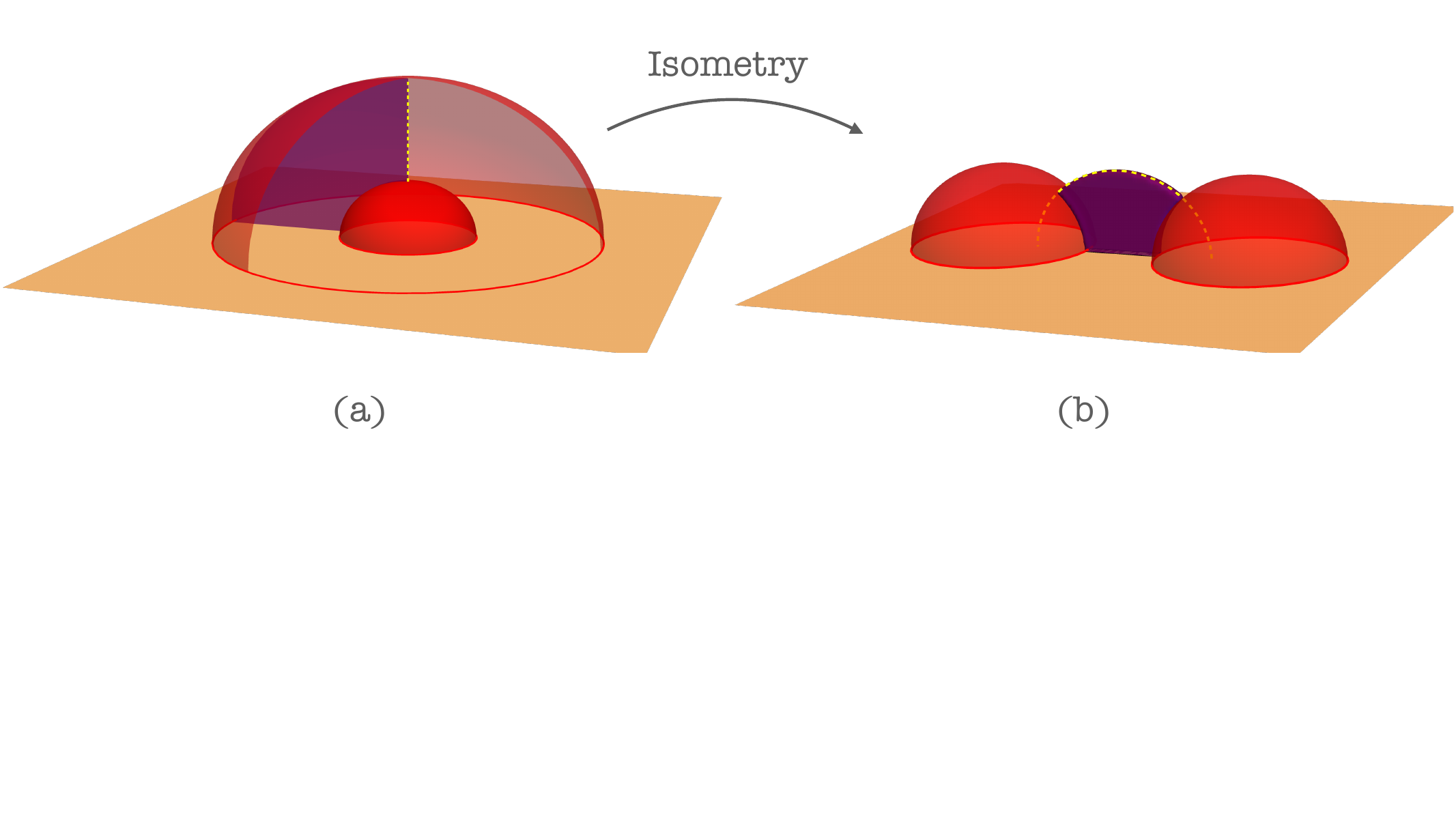}
                    \caption{The two figures are equivalent under the Mobius transformation. When changing the boundary function by $f(z) \rightarrow \frac{af_0(z)+b}{cf_0(z)+d}$ where $f_0(z)$ is also a solution for schwarzian equation for two-point function. This transformation changes the bulk too because of definition of Roberts map \ref{eq:43}}
                    \label{fig:isometry}
                \end{figure}

                At the conformal boundary which is $y=0$ we have via roberts map $Z=f(z)$. The $f(z)$ has branch points through which the branch cuts run. These solutions are specific to the case of Euclidean $AdS_3$ and are only unique upto isometry of Euclidean $AdS_3$ itself. 
                
                Euclidean $AdS_3$ which is the hyperbolic space of 3 dimesnions can be mapped to 3-dimensional disk (a ball) whose conformal boundary is a 2-sphere which in this case is a Riemann sphere because it has $\infty$ well defined on it. The global conformal transformation defined on the Riemann sphere generated by $L_{-1}$, $L_{0}$, $L_{1}$ which are generators of translations and scaling which together form a representation of mobius transformation. This mobius transformation can be continued into the bulk and forms an orientation preserving symmetry of Euclidean $AdS_3$. 

                So out solution $f(z)$ of schwarzian equation can be extended family of solutions based on mobius transformation

                \begin{equation*}
                    f(z) \rightarrow \frac{af_0(z)+b}{cf_0(z)+d}
                \end{equation*}

                where $f_0(z)$ is another known solution.

                After the isometric transformation the branch cut becomes a connection running between the domes. The black hole horizon also changes in the new geometry and from running through the spine of both domes now forms a geodesic between the two domes represented in Figure \ref{fig:isometry} with dotted yellow lines.

\chapter{Three-point functions}

    \section{Three point correlation functions of light operators}
    We review here the three-point function\cite{Freedman:1998tz} \cite{ammon2015gauge} in AdS/CFT.

    We borrow the argument of bulk-to-boundary propagator from \ref{2-point-function} but instead of two propagator meeting at a single point, we now impose the condition that all three propagator with different points on boundary meet at same point in the bulk. 

    The three-point amplitude is defined by combination of the propagators:
    
    \begin{equation}
        A(\vec{x}, \vec{y}, \vec{z}) 
    = - \int \frac{d^{d}w dw_0}{w_0^{d+1}}
    K_{\Delta_1}(w,\vec{x}) K_{\Delta_2}(w,\vec{y}) K_{\Delta_3}(w,\vec{z})
    \end{equation}

    By translational invariance we can set $\vec{z} = 0$ and using inversion $x_{i}=\frac{x'_{i}}{{x'^2}}$ giving us the three-point amplitude as
    
    \begin{equation}
        A(\vec{x},\vec{y},0) =-
    \frac{1}{|\vec{x}|^{2 \Delta_1}}
    \frac{1}{|\vec{y}|^{2 \Delta_2}}
    \frac{\Gamma(\Delta_3)}{\pi^{\frac{d}{2}} \Gamma(\Delta_3-  \frac{d}{2})} 
    \int \frac{d^{d}w' \, dw'_0}{(w'_0)^{d+1}} \, K_{\Delta_1}(w', \vec{x'})
    K_{\Delta_2}(w', \vec{y'}) \,(w'_0)^{\Delta_3}
    \end{equation}

    The integral has the form
    \begin{equation}
        \int_0^\infty dz_0 \int d^d\vec z frac{z_0^a}{ [z_0^2+(\vec z - 
    \vec x)^2]^b[z_0^2+(\vec z-\vec y)^2]^c}  = 
    I[a,b,c,d] |\vec x - \vec y |^{{1}+{a}+d-2b-2c  }
    \end{equation}

    where 
    \begin{equation}
         I[a,b,c,d]=
    \frac{\pi^{d/2}}{2} \frac{\Gamma[{\frac{a}{2}}+{\frac{1}{2}}]
    \Gamma[b+c-{\frac{d}{2}}-{\frac{a}{2}}-{\frac{1}{2}}]   }{ \Gamma[b]\Gamma[c]}
    \\ 
    \frac{\Gamma[{\frac{1}{2}}+{\frac{a}{2}}+{\frac{d}{2}- 
    b]}{\Gamma[{\frac{1}{2}}
    +{\frac{a}{2}}+{\frac{d}{2}}-c]}}{{\Gamma[1+a+d-b-c] }}
    \end{equation}

    So out three-point amplitude after integrating has the form of where 

    \begin{equation}
         \frac{1}{|\vec{x}|^{2 \Delta_1}
    |\vec{y}|^{2 \Delta_2}
    |\vec{x'}-\vec{y'}|^{(\Delta_1 + \Delta_2 - \Delta_3)} }
    =
    \frac{1}{|\vec{x}|^{ \Delta_1+ \Delta_3 - \Delta_2}
    |\vec{y}|^{ \Delta_2 + \Delta_3 - \Delta_1}
    |\vec{x}-\vec{y}|^{(\Delta_1 + \Delta_2 - \Delta_3)} }
    \end{equation}

    Now we can translate it again by $\vec x 
    \rightarrow ( \vec x- \vec z)$, $\vec y \rightarrow (\vec y- \vec z)$ and the above term becomes
    
    \begin{equation}
    \frac{1}{|\vec{x}-\vec{y}|^{ \Delta_1+ \Delta_3 - \Delta_2}
    |\vec{y}-\vec{z}|^{ \Delta_2 + \Delta_3 - \Delta_1}
    |\vec{z}-\vec{x}|^{(\Delta_1 + \Delta_2 - \Delta_3)} } \equiv f(\vec{x},\vec{y},\vec{z})
    \end{equation}

    This is of the form of three-point function of CFT.
    
    \section{Heavy-Heavy-Heavy correlation function}
        
        We have in earlier chapter described how two-point functions for heavy operators are described from the gravitational side in arbitrary dimensions. This is far harder for three-point function but we exploit the existence of Banados metric in three-dimensions along with existence of Roberts map. So we can attempt to find the three point function of heavy operators in three dimensions.

        Using the Banados metric from earlier we put in the boundary stress tensor for three operator insertions in z-plane \ref{eq:41} which gives us.
        
            \begin{equation}
            \langle T (z) \rangle = \frac{\langle T (z) \bigO_{\Delta_1} \bigO_{\Delta_2} \bigO_{\Delta_3} \rangle}{\langle  \bigO_{\Delta_1} \bigO_{\Delta_2} \bigO_{\Delta_3} \rangle}
            \end{equation} 

        which after applying ward identity and the form of three-point function \footnote{The form of three point correlation function is \begin{equation}
            \langle\bigO_{\Delta_1} \bigO_{\Delta_2} \bigO_{\Delta_3} \rangle = C_{ijk} (z_1-z_2)^{\Delta_3-\Delta_1-\Delta_2} (z_2-z_3)^{\Delta_1-\Delta_2-\Delta_3} (z_1-z_3)^{\Delta_2-\Delta_1-\Delta_3}
        \end{equation} where \(C_{ijk}\) is structure constant, this doesn't appear in stress-tensor because of ward identity it cancels. A reminder for ward identity is
        \begin{equation}
            \left\langle T(z) \bigO_{\Delta_1} \bigO_{\Delta_2} \bigO_{\Delta_3} \right\rangle = \sum_{i=1}^3\left(\frac{\Delta_i}{(z-z_i)^2} +\frac{1}{z-z_i}\frac{\partial}{\partial z_i}\right)\left\langle \bigO_{\Delta_1} \bigO_{\Delta_2} \bigO_{\Delta_3} \right\rangle
        \end{equation}}

    \vspace{2pt}
    
        results in \footnote{$\langle T(z) \rangle = -\frac{c}{6} L(z) $}

        \begin{equation}\label{eq:2.1}
            L(z) = - \frac{1}{4 (z-z_1)(z-z_2)(z-z_3)} \sum_{i} \frac{M_i \prod_{j \neq i} (z_i-z_j)}{(z-z_i)}
        \end{equation}

        We insert the operators at 0,1 and $\infty$ which is a common practice for sake of convenience. 
        We get 
        
        \begin{equation}
       L(z) = - \frac{1}{4 z(z-1)} \left [ -\frac{M_1}{z} + M_2 + \frac{M_3}{z-1} \right]
        \end{equation}

        We need to now figure out the solution to Schwarzian differential equation as mentioned in \ref{eq:schwarzian}. The stress-tensor has three parameters which in theory of schwarz triangles leads us to no independent accessory parameters. We can write $f(z) = \frac{u_i}{u_j}$ where $u_i$ and $u_j$ separately solve the hyper-geometric differential equation. This leads us to schwarz triangle function \cite{harmer2007note}. 

        This for our case is:

        \begin{equation}\label{eq:2.2}
            f_{3pt} = iNz^{iR_1} \frac{_2F_1 \left( \frac{1}{2}[1 + i (R_1-R_2-R_3)], \frac{1}{2}[1 + i (R_1+R_2-R_3)],1+iR_1;z\right)}{_2F_1 \left( \frac{1}{2}[1 - i (R_1-R_2+R_3)], \frac{1}{2}[1 - i (R_1-R_2+R_3)],1-iR_1;z\right)}
        \end{equation}

        Where $R_i$ similarly from the two-point function argument is \(R_i=\sqrt{M_i-1}\) which is the horizon of radius of black-hole at that particular insertion point. This solution has branch points at 0,1 and $\infty$ which is our point of operator insertion. As theory of complex function suggests now we can choose our branch cut running from these branching points.

        \begin{figure}
            \centering
            \includegraphics[scale=0.6]{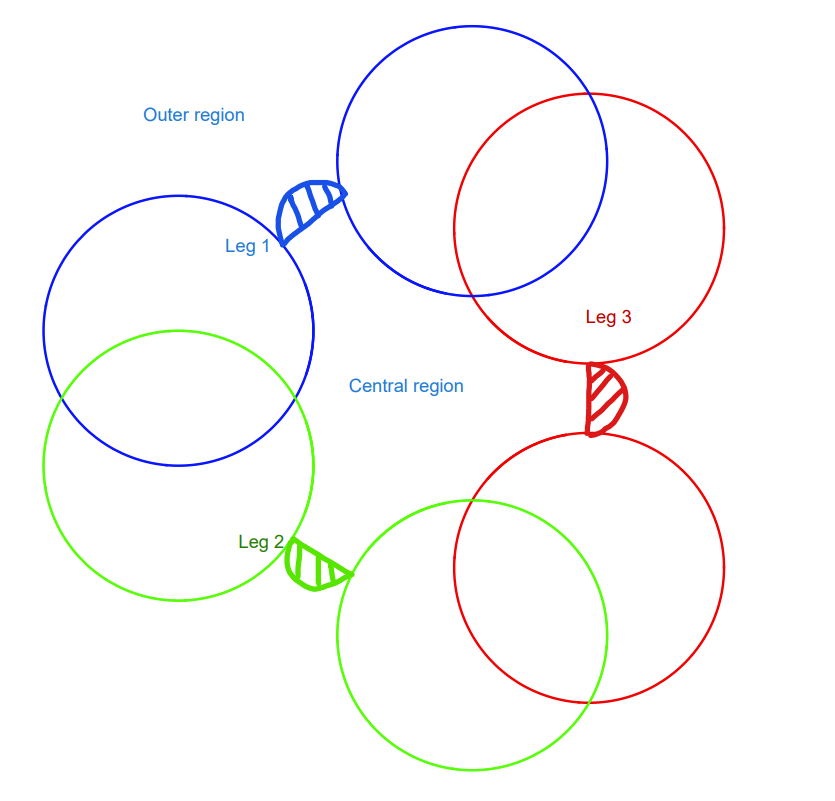}
            \caption{The three-point function geometry is represented here where the same colored circles are identified with each other and there is a branch cut running along then extended in the bulk just like that in two-point function. Each identification when continued into bulk becomes a dome. This is a isometry of genus-2 representation of schottky group \cite{krasnov2000holography} with added branch cuts. Each branch cut represents a pair heavy operator insertion as we have seen in two-point function case.}
            \label{fig:three-point identification}
        \end{figure}

        \subsubsection{Euclidean wormhole interpretation}

        The geometry represented in Figure \ref{fig:three-point identification} forms two surfaces. One is an inner region inside the three branch cuts and one is out side. This existence of two disconnected set on the boundary can now be interpreted as a wormhole \cite{susskind2005wormholes}. We see that we get three branch cut. In the isometry of two-point function we saw that there is a branch cut when we insert two heavy operators. Here we insert three operators but we get a three branched regions disconnecting the inside region from outside. The three branched region can now be interpreted as six operators being inserted but we should have expected only three because we insert three operators. This leads to a Euclidean wormhole with two asymptotic region with each region seen as having three insertions. The central region can be now thought of as the mouth of a wormhole. We now describe it with Maldacena-Maoz \cite{Maldacena_2004} wormhole metric:

        \begin{equation}\label{eq:2.3}
            ds^2=d\rho^2+cosh^2\rho \ d\Sigma^2
        \end{equation}

        where the metric $d\Sigma^2$ has constant negative curvature. The asymptotic region of $AdS_3$ Black-Hole are reached with $\rho \rightarrow \pm \infty$. 

    \subsection{Action}

    We now evaluate the action for the three-point function. We first review the liouville theory and apply it to the metric so we can exploit their property and evaluate the Integral much faster.

        \subsubsection{Liouville theory}

        The Maldacena-Maoz metric is solution to Einstein's field equation in three-dimensions for negative curvature. For empty AdS with no insertions the metric becomes

        \begin{equation}\label{eq:2.4}
            y = Y = \frac{w_1}{cosh \rho} \ ; \ z = Z = w_1 tanh\rho + iw_2
        \end{equation}

        Where 'w' is a new complex variable introduces with $w=w_1+iw_2$

        and so we get 

        \begin{equation}\label{eq:2.5}
            ds^2=d\rho^2+cosh^2\rho \ \frac{4 dw d\bar{w}}{(w+\bar{w})^2}
        \end{equation}

        In this metric we reach the boundary when $\rho \rightarrow \infty$ which in Poincare metric happens with $y=Y=0$. In this new metric we can also reach the boundary by taking $w_1=0$ for any $\rho$ because in the above metric it is the denominator. A direct consequence of this observation is that the asymptotic boundary is along the imaginary axis where $Re(z)=Re(Z)=0$ where $w_1=0$. This leads us to realization that there is a single boundary and is generally true went $\Sigma$ is non-compact. We now see for compact $\Sigma$ we get two seperate asymptotic regions which will we useful for our case of three-point function.

        We now for $d=2$ take a line element such that it satisfies the classical liouville equation \cite{zamolodchikov2007lectures} (which is the einstein's equation in 2d) for a field $\phi(w,\bar{w})$ for negative cosmological constant. \footnote{It is defined so because the Ricci scalar $$R=-4 e^{-\phi(w,\bar{w})}\frac{\partial^2 \phi(w,\bar{w})}{\partial w\partial\bar{w}}$$ and $R=-2$ for the 2D boundary surface}

        \begin{equation}\label{eq:2.6}
            d\Sigma^2= e^{\phi(w,\bar{w})} dw d\bar{w} \  ;  \  \frac{\partial^2 \phi(w,\bar{w})}{\partial w\partial\bar{w}}  = \frac{1}{2} e^{\phi(w,\bar{w})}
        \end{equation} 

        Now we can exploit the fact that we only need to find the Liouville solution. First we do a Fefferman-Graham expansion of the function $\rho(y,z,\bar{z})$, $w(y,z,\bar{z})$ and $\bar{w}(y,z,\bar{z})$ to do local change of variables such that for small $y$ the Maldacena-Maoz metric becomes like that of Banados metric.

            \begin{equation}\label{eq:2.7}
                \rho = - \log(y/2)+(1/2)\phi(z,\bar{z})+\frac{y^2}{4}\partial\phi(z,\bar{z})\bar{\partial}\phi(z,\bar{z}) + \dots
            \end{equation}

            \begin{equation}\label{eq:2.8}
                w = z-\frac{y^2}{2}\partial\phi(z,\bar{z}) + \dots \ ; \ \bar{w} = \bar{z}-\bar{\partial}\phi(z,\bar{z})
            \end{equation}

            Putting in the metric we read off the stress-tensor component from Banados metric as

            \begin{align}\label{eq:2.9}
                L(z) = 1/4 (\partial \phi (z,\bar{z}))^2 - 1/2 (\partial^2 \phi (z,\bar{z})) &&
                \bar{L}(\bar{z}) = 1/4 (\bar{\partial} \phi (z,\bar{z}))^2 - 1/2 (\bar{\partial}^2 \phi (z,\bar{z}))
            \end{align}

            Both of the stress-tensor are holomorphic (or subsequently anti-holomorphic) because $\partial \bar{L}(\bar{z})=\bar{\partial} {L}(z)=0$ after using the liouville equation of motion. 

            The generic solution for Liouville equation of motion in terms of holomorphic and anti-hol0morphic function is given by

            \begin{equation}
                \phi(z,\bar{z})= \log\left(\frac{4 \partial f \overline{\partial f}}{(f+f)^2}\right)
            \end{equation}
            
            When this general solution is put up in the equation \ref{eq:2.9} we recover the schwarzian derivative solution. $L(z) = -\{f,z\}$ and so for c.c.
        
            So these holomorphic functions f can be used to derive the Liouville field $\phi$.

            We see for example the liouville field of 2pt black hole we get after substituting $f_{BH}$ at 0 and $\infty$

            \begin{equation}\label{eq:2.10}
                e^{\phi_{2pt}} = \frac{R_h^2}{w\bar{w}cos^2\left(\frac{R_h}{2} \ln(w\bar{w})\right)}
            \end{equation}

            This function diverges when denominator vanishes which is at $w\bar{w}=e^{n \pi R_h}$ for integral value of n. This divergence corresponds when the metric reaches the asymptotic boundary. At the boundary from FG expansion $w=z$ and then all the literature of dome and branch cut is re-introduced to describe the singularity of $\phi_{BH}$.

        \subsubsection{Setting up the Action}

        Recall from the discussion of two-point functions of banana geometry in (d+1) dimension. We evaluated a d-dimensional integral with bulk boundary and stretched horizon term with counter-terms. Here we go ahead to do so for 2-d.

        The gravitation action becomes in this case:

        \begin{equation}\label{eq:2.11}
            8 \pi G_N S_{bulk+boundary+ct}= -\frac{1}{2}\int_{M} d\rho d^2z \sqrt{g}(R-\Lambda) + \int_{\partial M} d^2z \sqrt{h}K+ \int_{\partial M} d^2z \sqrt{h}
        \end{equation}
        We set for $Ads_3$ as earlier discussed $L_{AdS}=1$, $R=-6$, $\Lambda=-2$ and extrinsic curvature up-to leading term as $K=-2$. What remains is to figure out the metric of bulk and boundary for various sub-parts. I total combined we write this as 

        \begin{equation}
            4 \pi G_N S =\left(V-\frac{1}{2}A\right)
        \end{equation}

            Which represents the bulk volume and area of asymptotic boundary. We also need to add the stretched horizon term on top which makes the total action as

        \begin{equation}
            4 \pi G_N S = \left(V-\frac{1}{2}A\right)+ \frac{1}{2} \int_{streched \ horizon} d^2z \sqrt{h}K
        \end{equation}

        Now we'll patiently evaluate the term for central region which is the throat of the wormhole, and the region accessed by each set of operator insertion which are isometries of two-point function, and finally the stretched horizon. But first we talk about cut-off terms which isn't trivial in this case.

        If we choose our cut-off surface in a straightforward manner by looking at the FG expansion of $\rho$ in Eq \ref{eq:2.7} where boundary is $y \rightarrow 0$ or equivalently $\rho \rightarrow \pm \infty$. We can say that cut-off is surface of constant rho such that

        \begin{equation}
            \rho = \pm \ln \left ( \frac{\epsilon}{2} \right)
        \end{equation}

        but a subtlety to notice is that it should be of Poincare metric form in leading order to be meaningful cut-off surface 

        \begin{equation}
            ds^2 = \frac{dzd\bar{z}}{\epsilon^2} + \dots
        \end{equation}

        while on the other hand the cut-off surface we chose will give have metric of the form \footnote{We just put the cutoff \[\rho = \pm \ln \left ( \frac{\epsilon}{2} \right)\] in the wormhole metric and expand in epsilon. So \[Cosh^2 \left (\ln \left ( \frac{\epsilon}{2} \right) \right) = \frac{\left(\epsilon ^2+4\right)^2}{16 \epsilon ^2} \]}

        \begin{equation}
            ds^2=\frac{\left(\epsilon ^2+4\right)^2}{16 \epsilon ^2} e^{\phi(z,\bar{z})} dz d\bar{z}
        \end{equation}

        To find the correct cut-off surface we have to take another term from the FG expansion which gives the cut-off surface a $\phi$ dependence.

        So the correct cut-off surface is 

        \begin{equation}
            \rho = \pm \ln \left ( \frac{\epsilon}{2} \right) \pm \frac{\phi (z,\bar{z})}{2}
        \end{equation}

        Notice that $\phi (z,\bar{z})$ has singularities at point of insertion so when we deal with two-point function action we'll have to regulate for it carefully.

        \subsubsection{The central region action}

        We first evaluate the action of the central region which is bordered by all the branch cuts. 

        We go ahead and evaluate the volume of the bulk where the metric is the maldacena-maoz metric like mentioned above in \ref{eq:2.11} and keep in mind the cutoff surface described just above. We also use the Liouville equation of motion $\partial\bar{\partial} \phi = (1/2)*e^{\phi}$ to get the right form of the integral 

        We get the volume integral to be \footnote{We just integrate over limit \[ \rho = \ln \left ( \frac{\epsilon}{2} \right) \pm \frac{\phi (z,\bar{z})}{2}\] from $-\rho$ to $\rho$. We then get \[\frac{1}{\epsilon^2}+e^{\phi} \log(\frac{2}{\epsilon^2})-\frac{\phi e^{\phi}}{2}\] then apply liouville eom and apply product rule of derivatives to get the form \[\frac{\phi e^{\phi}}{2} = \bar{\partial}(\phi\partial\phi)-\partial\phi\bar{\partial}\phi\]}

        \begin{equation}\label{eq:2.12}
            V = \int d^2z \left(\frac{1}{\epsilon^2}+e^{\phi}*\ln\frac{2}{\epsilon} -\bar{\partial}(\phi\partial\phi)+\partial\phi\bar{\partial}\phi \right)
        \end{equation}

        Next is the area integral where the boundary metric is that of Maldacena-Maoz with $\rho=\text{cut-off surface}$

        \begin{equation}\label{eq:2.13}
            A = \int d^2z \left(\frac{2}{\epsilon^2}+e^{\phi}+\partial\phi\bar{\partial}\phi\right)
        \end{equation}

        We use the fact that \[\int d^2z e^{\phi} = 2\pi\] this is because the central region is a hyperbolic octagon with identification. A nice way to visualize is that this structure is a genus-2 surface with branch cuts. The central region is inside the branched regions so we're on the first sheet. So the central-region is a pure genus-2 surface. Now genus-2 surface has 4 cycles. 2 contractible and 2 non-contractible cycles. So we end up with 4 cycles, if we want to represent it with a polygon with identification we'll end up with a octagon with 4 edges identified with other 4 in a particular way. \footnote{A nice illustration is in a video \href{https://www.youtube.com/watch?v=G1yyfPShgqw}{here}}. The area of hyperbolic polygon is $A=(n-2)\pi - \sum_{i} \theta_{i}$ \cite{HyperbolicArea}. Area for a right angled hyperbolic octagon turn out be a nice $2\pi$.

        Combining everything leads to a central region action of:

        \begin{equation}\label{eq:2.14}
            8\pi G_N S_{central} = \int_{W} d^2z (\partial\phi\bar{\partial}\phi + e^{\phi}) +  \iota \int_{\partial W} dz \phi\partial\phi - \frac{1}{2G_N} (1-\ln(2/\epsilon))
        \end{equation}

         For second term we used the stokes theorem to take it to boundary, the third term comes by evaluating the area of hyperbolic octagon.

         \subsubsection{Leg Region action}

            The central region action is now evaluated which has no singularities in $\phi$ but we now have to evaluate for the set of two-point function which is accessed when we cross the three branch cuts. These leg regions have contribution of six times rather than three because of wormhole interpretation. These operator insertions have singularities which affect the earlier $\phi$. To avoid that we go to a new set of co-ordinates and change the metric of wormhole accordingly to use for the calculation of action.

            The new co-ordinate which we map the Poincare boundary to is defined using a common euclidean AdS coordinate:

            \begin{align} \label{eq:2.15}
                Y=e^{\tau}sech(\Tilde{\rho})  &&
                Z=e^{\tau + \iota \theta}tanh(\Tilde{\rho})  &&
                \bar{Z}=e^{\tau - \iota \theta}tanh(\Tilde{\rho})
            \end{align}

            The metric in this case becomes \footnote{\[2\tau = f_i + \bar{f}_i \] and \[2\iota \theta = f_i - \bar{f}_i\] where f and $\bar{f}$ are arbitary holomorphic and anit-holomorphic function and not necessarily same as that of central region.}

            \begin{equation} \label{eq:2.16}
                ds^2 = d\Tilde{\rho}^2 + cosh^2(\Tilde{\rho})d\tau^2+sinh^2(\Tilde{\rho}) d\theta^2
            \end{equation}

            The cutoff will be surface of constant $\Tilde{\rho}$ which will be cone surfaces something we say in Two-point function discussion in earlier chapter. 

            The cut-off surface like for central region is defined as:

            \begin{equation} \label{eq:2.17}
                \Tilde{\rho} = -\ln \left ( \frac{\epsilon}{2} \right) \pm \frac{\Tilde{\phi}(z,\bar{z})}{2}
            \end{equation}

            Where the $\Tilde{\phi}$ doesn't have singularities and they don't show up in the metric so they follow the Laplace equation of motion $\bar{\partial}\partial \phi = 0$. This can be thought also as that we're not in the throat of wormhole anymore and looking it from the outside where we can only observe two operator insertions and their correlation geometry. The solution to generic laplace equation is $\Tilde{\phi}= \ln (\partial f_i \bar{\partial f_i})$

            This gives the metric of cut-off surface as wanted that of $\frac{dz d\bar{z}}{\epsilon^2}$.

            We bring attention to one subtlety. The $\Tilde{\phi}$ and $\phi$ along with its derivatives do need to match at boundary of central region and leg region. This is our gluing condition for smooth and continuous cutoff surface across the whole region.

            The volume integral is after integrating over $\rho$ over the newly talked about metric and then changing the measure from $(\tau,\theta) \rightarrow (z,\bar{z})$ i.e $d\tau d\theta = e^{\Tilde{\phi}} dz d\bar{z}$

            \begin{equation*}
                V = \int d^2z \ \frac{1}{2 \epsilon^2} - \frac{e^{\Tilde{\phi}}}{4}
            \end{equation*}

            and area integral defined over constant cut-off surface

            \begin{equation*}
                A = \int d^2z \ \frac{1}{ \epsilon^2} + \frac{\partial\Tilde{\phi}\bar{\partial}\Tilde{\phi}}{2}
            \end{equation*}

            Combining in the same way of $\frac{1}{4\pi G_N} \left(V-\frac{1}{2}A\right)$

            We get action of each leg after using product rule, stokes theorem and applying Laplace equation of motion $\bar{\partial}\partial \phi = 0$ we get

            \begin{equation} \label{eq:2.18}
                8\pi G_N S_{Leg} = \frac{1}{2}\int_{L_i} d^2z (\partial\Tilde{\phi}\bar{\partial}\Tilde{\phi} - e^{\Tilde{\phi}}) - \iota \int_{\partial L_i} dz  \Tilde{\phi}\partial\Tilde{\phi}
            \end{equation}

            Here we notice that it is of the same form as that of central region action in Equation \ref{eq:2.14} but with a negative sign on $e^{\Tilde{\phi}}$ (which is the Liouville cosmological term). This is magically fixed by the streched horizon action which has been essential in out discussions so far.

            \subsubsection{Stretched horizon action}
            
                We take the cut-off surface here to be ones with $\Tilde{\rho}=\epsilon$ which has been our recipe till now even in two-point function section. We had seen that we take the surface just epsilon above the horizon. The induced metric is then $h=\epsilon^2$. The integral in the $(z,\bar{z})$ plane becomes 
                
                \begin{equation}
                    \int_{Stretch} d^2z \ \epsilon \ K e^{\Tilde{\phi}}
                \end{equation}

                The extrinsic curvature in leading order in taken as $K = g^{ab}K_{ab} = \frac{1}{\epsilon} + \dots$. Here we assume that for central region there is no stretched horizon but rather only for leg terms. This has to do with simplicity arguments and regardless the final answer matches. An interpretation for this simplicity is that the stretched horizon 

                So for each leg region we have 

                \begin{equation}\label{eq:2.19}
                    8\pi G_N S_{stretch} = \int  d^2z e^{\Tilde{\phi}}  \sqrt{h}K = \int_{L}  d^2z e^{\Tilde{\phi}}
                \end{equation}

            \subsubsection{The Total Action}

                We combine equations \ref{eq:2.14}, \ref{eq:2.18} and \ref{eq:2.19} to get the complete action which is

                \begin{equation}
                S = S_{central} +\sum_{i} I_{L_i} + I_{L_i-GHY,str}
                \end{equation}

                Which is 
                \begin{multline} \label{eq:2.20}
                    8 \pi G_N S = \int_{W} d^2z (\partial\phi\bar{\partial}\phi + e^{\phi}) + \sum_{i} \frac{1}{2}\int_{L_i} d^2z (\partial\Tilde{\phi_i}\bar{\partial}\Tilde{\phi_i} + e^{\Tilde{\phi_i}}) \\ 
                    -  \iota \int_{\partial W} dz \phi\partial\phi - \sum_{i}  \iota \int_{\partial L_i} dz \Tilde{\phi_i}\partial\Tilde{\phi_i} \\
                   - \frac{1}{2G_N} (1-\ln(2/\epsilon))
                \end{multline}            

                Now these to compare we see from Equation (17) of \cite{HADASZ2004493} that they define a regularized Liouville action ($S_L$) for three hyperbolic singularities. It also matches the classical DOZZ formula in the limit $b \rightarrow 0$. The log of the three point function \[ \log \Tilde{C} = - \frac{1}{b^2} S_L[\phi] + \text{const}\] 
                $\Tilde{C}$ is the symmetric three-point function. For large central charge c , $c=1+6(b+b^{-1})^2$ so for classical limit $b \rightarrow 0$, $\frac{1}{b^2}=\frac{c}{6}$. 

                The action in Eq (17) \cite{HADASZ2004493} 
                \begin{multline}\label{eq:2.21}
                S_L[\phi] = \frac{1}{2 \pi}\int_{W} d^2 z
                \left(\partial\phi \bar{\partial}\phi + e^{\phi}\right)  + \frac{1}{2 \pi} \sum_{i=1}^3\left(\int_{{L}_i} d^2z\left(\partial \tilde{\phi}_i \bar{\partial} \tilde{\phi}_i + e^{\tilde{\phi}_i}\right)  \right) \\
                - \frac{i}{\pi} \int_{|z|=1/\epsilon} dz \ \phi \partial \phi - \frac{i}{\pi} \sum_{i=1}^3  \int_{|z-z_i|=\epsilon}dz \ \tilde{\phi}_i \partial\tilde{\phi}_i  - \left(4 + \sum_{i=1}^3(1 - R_i^2)\right) \ln(\epsilon)\,.
                \end{multline}

                The equations \ref{eq:2.21} matches \ref{eq:2.20} quite closely which gives the relation between the gravitational action we've analyzed and the liouville action.

                The relation is:

                \begin{equation}
                    S = \frac{c}{6}  S_L[\phi] -\frac{c}{3} (1-  \log2) + \frac{c}{6} \sum_{i=1}^3(1 - R_i^2) \ln(\epsilon)
                \end{equation}

                The last term is cancelled by counter-term that arise via two-point function. We'll do a short calculation of it in Appendix.

                \begin{equation}
                    S = \frac{c}{6}  S_L[\phi] -\frac{c}{3} (1- \\ln 2)
                \end{equation}

                After dealing with two-point function. This leads to the three-point function with singularities which has cosmological constant $\mu = \frac{1}{4 \pi b^2}$ in the classical limit of the DOZZ formula talked above.

                We see that the final saddle-point approximation of the path integral will lead to

                \begin{equation}
                    e^{-S} = |C_L(z_1,z_2,z_3)|^2
                \end{equation}

                Where $C_L$ is the classical limit of three-point function. This is reproduced in \cite{Chandra:2022bqq} which does this calculation for defects matching Eq. (4.47) and also shows that the gravitational path integral is not the three-point function but rather the squared average of it representing an ensemble which happens because of wormhole having multi-asymptotic boundary. 
        

    \chapter{Conclusion and future direction}

    We've explored the two-point and three-point function of Heavy operators. Two-point operator was explored in (d+1) dimensional geometry because of the description of new geometry "Space-time banana" and was computable for all dimensions. For three-point function we have much less analytical control and we fall to $Ads_3$ to explore it. In $Ads_3$ the gravity is topological and hence we have an exact metric \cite{Bañados:1999} which we exploited to solve the problem. In the three-point function the final correlator we get matches the "Square" of the DOZZ formula. This shows that the geometry describes the dual to ensemble averages of the CFT states. We wouldn't have a disconnected surface and be begging for an ensemble interpretation had this been a dual to a field theory.

    The $AdS_3$ has another vacuum solution other than black-holes which is the conical defect solution. These are also dual to the heavy operator insertion and the dome geometry is able to describe is as well by just analytically continuing out boundary function for $M > 1$ to that of $M<1$, this in the three-point function case makes the throat of the wormhole to collapse and make the domes meet at the middle with the angle describing the conical geometry. The three-point correlator for the defect geometry was explored in \cite{Chandra:2022bqq} where they also get the equivalent "square of DOZZ formula" with different factors.

    This work was a review of \cite{Vieira:2023bqv}\cite{Vieira:2023jye} which we have used as a warmup to explore the realm of Heavy operator geometry without any help from lagrangians. Our future direction is to compute the correlation between heavy and light operators which is under works. Like the Heavy-Heavy-Light configuration which on the field theory side has been described in \cite{Collier:2019weq}. The gravity side expectation is these are decaying black-holes in euclidean space. The light operator is a particle escaping from the black hole (heavy operator). Given that heavy operator in $Ads_3$ can be black-hole or conical defects gives us a wide range of possibilities to explore new physics using the new description of heavy correlators in this thesis.

        \chapter{Appendix}

        \subsubsection{The Two-point black hole action in $AdS_3$}

    We calculated this action when we were dealing with leg action region. 
    
    \begin{equation}
        16 \pi G_N S_{2pt} =  \int d^2 z \left(\partial \tilde{\phi} \bar{\partial} \tilde{\phi} + e^{\tilde{\phi}(z,\bar{z})}\right) - \frac{i}{2} \int dz \ \tilde{\phi} \partial \tilde{\phi}
    \end{equation}

    The Laplace solution corresponding to two hyperbolic singularities located at $z = 1$ and $z = -1$  is derived from the relation
    
    \begin{equation}
    \tilde{\phi}_{2pt}(z,\bar{z}) = \\ln\left(\frac{4 R_h^2}{|z^2 - 1|^2}\right)
    \end{equation}
    Inserting this into the action above, we find
    
    \begin{equation}
    S = \frac{1}{4 \pi G_N} \int d^2 z \left(-\frac{|z|^2}{|z^2 - 1|^2} + \frac{R_h^2}{|z^2 - 1|^2} \right)
    \end{equation}
    
    where we cut out a region of size $\epsilon$ around each of the insertions.  The $R_h$ dependent term gives
    
    \begin{equation}
    \frac{1}{4\pi G_N}\int d^2 z \frac{R_h^2}{|z^2-1|^2} = -\frac{1}{4 G_N} R_h^2 \\ln(\epsilon)
    \end{equation}
    
    So we see that
    
    \begin{equation}
    S = -\frac{1}{4 G_N} R_h^2 \ln(\epsilon) + C \label{act99}
    \end{equation}
    
    where the constant $C$ does not depend on $R_h$.

    To fix $C$, we note that when $R_h^2 = -1$, we should have the vacuum.
    The vacuum partition function can be calculated from the conformal anomaly \cite{Chandra:2022bqq}, we get the answer as
    
    \begin{equation}
    \\ln(Z(S^2)) = \frac{1}{2 G_N} \ln\left(\frac{R}{\epsilon}\right)
    \end{equation}
    
    So we have
    
    \begin{equation}
    S|_{R_h\rightarrow -1} = \frac{1}{4 G_N} \ln\epsilon + C = -\frac{1}{2 G_N} \\ln\left(\frac{R}{\epsilon}\right)
    \end{equation}
    
    which allows us to read off the value of the constant $C$,
    
    \begin{equation}
        C = -\frac{1}{2 G_N}\\ln R + \frac{1}{4 G_N} \ln\epsilon\
    \end{equation}
    
    Finally, we find the action \ref{act99} becomes
    
    \begin{equation}
        S = \frac{1}{4 G_N}(1 - R_h^2) \ln\epsilon -\frac{1}{2 G_N} \\ln R
     \end{equation}

    Hence, to leading order, the action is given by
    \begin{equation}
    S \simeq \frac{1}{4 G_N}(1-R_h^2)\ln\epsilon\,,
    \end{equation}
    and so for each operator, we must add a counter-term of the form
    
    \begin{equation}
        S_{ct}(R_h) = -\frac{1}{8 G_N}(1-R_h^2) \ln\epsilon
    \end{equation}

    We can see that
    
    \begin{equation}
        \frac{1}{4 G_N}(1-R_h^2) = 2\Delta
    \end{equation}
    so we can see that 
    \begin{equation}
        e^{-S} = \epsilon^{-2\Delta}
    \end{equation}
    which is the two point function for insertion of cut-out region of radius $\epsilon$. To get the correlation of \\
    $\langle \bigO(-1 \pm \epsilon) \bigO(1 \pm \epsilon)\rangle = (2(1 \pm \epsilon))^{-2\Delta}$ we'll need to add a constant shift to the integral which doesn't affect the leading order for counter-term.



    \bibliographystyle{plain}
    \bibliography{bibliography.bib}

\begin{thebibliography}{10}

\bibitem{Vieira:2023jye}
Jacob Abajian, Francesco Aprile, Robert~C. Myers, and Pedro Vieira.
\newblock {Holography and Correlation Functions of Huge Operators: Spacetime Bananas}.
\newblock {\em J. High Energ. Phys. 2023, 58 (2023)}.

\bibitem{Vieira:2023bqv}
Jacob Abajian, Francesco Aprile, Robert~C. Myers, and Pedro Vieira.
\newblock {Correlation Functions of Huge Operators in AdS$_3$/CFT$_2$: Domes, Doors and Book Pages}.
\newblock 2023.

\bibitem{ammon2015gauge}
M.~Ammon and J.~Erdmenger.
\newblock {\em Gauge/Gravity Duality}.
\newblock Cambridge University Press, 2015.

\bibitem{Balasubramanian:1999re}
Vijay Balasubramanian and Per Kraus.
\newblock {A Stress tensor for Anti-de Sitter gravity}.
\newblock {\em Commun. Math. Phys.}, 208:413--428, 1999.

\bibitem{Bañados:1999}
Máximo Bañados.
\newblock {Three-dimensional quantum geometry and black holes}.
\newblock {\em AIP Conference Proceedings}, 484(1):147--169, 07 1999.

\bibitem{PhysRevD.7.2333}
Jacob~D. Bekenstein.
\newblock Black holes and entropy.
\newblock {\em Phys. Rev. D}, 7:2333--2346, Apr 1973.

\bibitem{Cardy:2017qhl}
John Cardy, Alexander Maloney, and Henry Maxfield.
\newblock {A new handle on three-point coefficients: OPE asymptotics from genus two modular invariance}.
\newblock {\em JHEP}, 10:136, 2017.

\bibitem{Chandra:2022bqq}
Jeevan Chandra, Scott Collier, Thomas Hartman, and Alexander Maloney.
\newblock {Semiclassical 3D gravity as an average of large-c CFTs}.
\newblock {\em JHEP}, 12:069, 2022.

\bibitem{Maloney:2019weq}
Scott Collier, Alexander Maloney, Henry Maxfield, and Ioannis Tsiares.
\newblock {Universal dynamics of heavy operators in CFT$_{2}$}.
\newblock {\em JHEP}, 07:074, 2020.

\bibitem{Collier:2019weq}
Scott Collier, Alexander Maloney, Henry Maxfield, and Ioannis Tsiares.
\newblock {Universal dynamics of heavy operators in CFT$_{2}$}.
\newblock {\em JHEP}, 07:074, 2020.

\bibitem{DiFrancesco:1997nk}
P.~Di~Francesco, P.~Mathieu, and D.~Senechal.
\newblock {\em {Conformal Field Theory}}.
\newblock Graduate Texts in Contemporary Physics. Springer-Verlag, New York, 1997.

\bibitem{DORN1994375}
H.~Dorn and H.-J. Otto.
\newblock Two- and three-point functions in liouville theory.
\newblock {\em Nuclear Physics B}, 429(2):375--388, 1994.

\bibitem{Emparan:1999pm}
Roberto Emparan, Clifford~V. Johnson, and Robert~C. Myers.
\newblock {Surface terms as counterterms in the AdS / CFT correspondence}.
\newblock {\em Phys. Rev. D}, 60:104001, 1999.

\bibitem{Freedman:1998tz}
Daniel~Z. Freedman, Samir~D. Mathur, Alec Matusis, and Leonardo Rastelli.
\newblock {Correlation functions in the CFT(d) / AdS(d+1) correspondence}.
\newblock {\em Nucl. Phys. B}, 546:96--118, 1999.

\bibitem{HADASZ2004493}
Leszek Hadasz and Zbigniew Jaskólski.
\newblock Classical liouville action on the sphere with three hyperbolic singularities.
\newblock {\em Nuclear Physics B}, 694(3):493--508, 2004.

\bibitem{harmer2007note}
M.~Harmer.
\newblock Note on the schwarz triangle functions, 2007.

\bibitem{unknown-author-no-date}
https://stackexchange.com/users/12602893/a-v s.
\newblock {What is the stretched horizon of a black hole?}

\bibitem{kaplan-no-date}
Jared Kaplan.
\newblock {Lectures on AdS/CFT from the bottom up}.

\bibitem{krasnov2000holography}
Kirill Krasnov.
\newblock {Holography and Riemann surfaces}.
\newblock {\em Adv. Theor. Math. Phys.}, 4:929--979, 2000.

\bibitem{Louko_2000}
Jorma Louko, Donald Marolf, and Simon~F. Ross.
\newblock Geodesic propagators and black hole holography.
\newblock {\em Physical Review D}, 62(4), Jul 2000.

\bibitem{Maldacena_2004}
Juan Maldacena and Liat Maoz.
\newblock Wormholes in ads.
\newblock {\em Journal of High Energy Physics}, 2004(02):053, mar 2004.

\bibitem{McGreevy:2000cw}
John McGreevy, Leonard Susskind, and Nicolaos Toumbas.
\newblock {Invasion of the giant gravitons from Anti-de Sitter space}.
\newblock {\em JHEP}, 06:008, 2000.

\bibitem{PhysRevD.58.064011}
Maulik~K. Parikh and Frank Wilczek.
\newblock An action for black hole membranes.
\newblock {\em Phys. Rev. D}, 58:064011, Aug 1998.

\bibitem{PhysRevD.33.915}
Richard~H. Price and Kip~S. Thorne.
\newblock Membrane viewpoint on black holes: Properties and evolution of the stretched horizon.
\newblock {\em Phys. Rev. D}, 33:915--941, Feb 1986.

\bibitem{Roberts:2012aq}
Matthew~M. Roberts.
\newblock {Time evolution of entanglement entropy from a pulse}.
\newblock {\em JHEP}, 12:027, 2012.

\bibitem{susskind2005wormholes}
Leonard Susskind.
\newblock Wormholes and time travel? not likely, 2005.

\bibitem{Witten:1998qj}
Edward Witten.
\newblock {Anti-de Sitter space and holography}.
\newblock {\em Adv. Theor. Math. Phys.}, 2:253--291, 1998.

\bibitem{ZAMOLODCHIKOV1996577}
A.~Zamolodchikov and Al. Zamolodchikov.
\newblock Conformal bootstrap in liouville field theory.
\newblock {\em Nuclear Physics B}, 477(2):577--605, 1996.

\bibitem{zamolodchikov2007lectures}
AB~Zamolodchikov and Alexander Zamolodchikov.
\newblock Lectures on liouville theory and matrix models.
\newblock {\em Google Scholar}, 2007.

\bibitem{HyperbolicArea}
Sarah Zhang.
\newblock A roundabout introduction to hyperbolic area, 2019.

\bibitem{10.1093/oso/9780198834625.001.0001}
Jean Zinn-Justin.
\newblock {\em {Quantum Field Theory and Critical Phenomena: Fifth Edition}}.
\newblock Oxford University Press, 04 2021.

\end{thebibliography}
\end{document}